\documentclass[
 aip,
 reprint,
 jcp,
 nolongbibliography
]{revtex4-2}

\usepackage{graphicx}
\usepackage{bbm}
\usepackage{dcolumn}
\usepackage{amsmath,amssymb,bm}
\usepackage{mathtools}
\usepackage[bookmarks=false,breaklinks=true]{hyperref}
\usepackage{xcolor}
\usepackage{scalerel}
\usepackage[caption=false]{subfig}
\usepackage{accents}
\usepackage[normalem]{ulem}
\newcommand{\vect}[1]{\boldsymbol{\mathbf{#1}}}
\newcommand{\dbtilde}[1]{\accentset{\approx}{#1}}
\newcommand{\angstrom}{\text{\normalfont\AA\ }}
\newlength\bshft
\def\fakebold#1{\ThisStyle{\ooalign{$\SavedStyle#1$\cr%
  \kern-\bshft$\SavedStyle#1$\cr%
  \kern\bshft$\SavedStyle#1$}}}

\hypersetup{
  pdfstartview = FitH,
  pdftoolbar   = false,
  pdfmenubar   = true,
  colorlinks   = true,
  linkcolor    = blue!50!black,
  urlcolor     = blue!50!black,
  citecolor    = blue!50!black
}

\begin{document}

\preprint{AIP/123-QED}

\title{Configurational entropy, transition rates, and optimal interactions for
rapid folding in coarse-grained model proteins}
\author{Margarita~Colberg}
\email{margarita.gladkikh@mail.utoronto.ca}
\author{Jeremy~Schofield}
\email{jeremy.schofield@utoronto.ca}
\affiliation{Chemical Physics Theory Group, Department of Chemistry, University
of Toronto, Toronto, Ontario, M5S 3H6, Canada}

\date{\today}

\begin{abstract}
Under certain conditions, the dynamics of coarse-grained models of solvated
proteins can be described using a Markov state model, which tracks the evolution
of populations of configurations. The transition rates among states that appear
in the Markov model can be determined by computing the relative entropy of
states and their mean first passage times. In this paper, we present an adaptive
method to evaluate the configurational entropy and the mean first passage times
for linear chain models with discontinuous potentials. The approach is based on
event-driven dynamical sampling in a massively parallel architecture. Using the
fact that the transition rate matrix can be calculated for any choice of
interaction energies at any temperature, it is demonstrated how each state's
energy can be chosen such that the average time to transition between any two
states is minimized. The methods are used to analyze the optimization of the
folding process of two protein systems: the crambin protein, and a model with
frustration and misfolding. It is shown that the folding pathways for both
systems are comprised of two regimes: first, the rapid establishment of local
bonds, followed by the subsequent formation of more distant contacts. The state
energies that lead to the most rapid folding encourage multiple pathways, and
they either penalize folding pathways through kinetic traps by raising the
energies of trapping states, or establish an escape route from the trapping
states by lowering free energy barriers to other states that rapidly reach the
native state.
\end{abstract}

\maketitle

\section{Introduction} \label{sec:intro}

Coarse-grained models of polymers\cite{Joshi:2021} and
proteins\cite{Kmiecik:2016,Kmiecik:2018} are designed to bridge the gap in time
scale between the motion of molecular components and slower, large-scale
structural changes. A wide variety of such models exist, including G\={o}
lattice models, in which monomers are restricted to lattice
sites\cite{Taketomi:1975,Ueda:1978,Go:1979,Go:1981,Abe:1981,Takada:2019},
elastic networks\cite{Tirion:1996,Yang:2009,Schofield:2012}, and off-lattice
linear chain models, which make use of a continuous force field that may include
quantum effects in an approximate way\cite{Orozco:2014}. These models have been
used to probe the mechanism of protein folding, the process by which a denatured
protein reaches its experimentally determined native
structure\cite{Larsen:2011}. The dynamics derived from coarse-grained models
indicate that short, fast-folding proteins follow a hierarchical folding
process. In this process, the backbone adopts secondary structural elements and
a small number of nonlocal contacts early on, and then subsequently folds in a
directed fashion along a dominant
pathway\cite{Karplus:1994/650,Fersht:1997/3,Dill:2008/289}. Other simulation
work has suggested that the folding is fast and efficient when the system is
free of bottlenecks or kinetic traps, and when multiple pathways exist to the
final state\cite{Wolynes:1995/1619,Onuchic:1992}. Experimental evidence,
primarily through studies of {\it cytochrome
c}\cite{Krishna:2004,Krishna:2006,Maity:2005} and RNase\cite{Hu:2016}, support
the picture that the initial phase of the folding process consists of the
formation of ``foldon" subunits made up of secondary
structures\cite{Bai:1995,Hu:2013,Hu:2016}.

In recent years, there has been an interest in engineering polymer and protein
systems that possess properties similar to those of naturally occurring
proteins\cite{Baker:2003,Richter:2013,Chen:2019,Zhou:2020}. Although a number of
design considerations have been identified that are associated with secondary
and tertiary structure, such as the use of residue sequences that have a
propensity to form $\alpha$-helices and other structural elements or that have
hydrophobic side chains to facilitate packing, the relative importance of each
design feature for a particular folded structure is not easily
determined\cite{Cordes:1996}. Experimental and simulation studies suggest that
secondary structure is important in providing the building blocks for foldons
that nucleate the folding process\cite{Dill:2008/289,Bai:1995,Hu:2013,Hu:2016},
but other types of structures might provide a similar framework if they satisfy
a set of physical characteristics. Understanding such requirements could provide
insight into how to design synthetic polymer systems with protein-like structure
and functionality.

Much recent work to connect the primary sequence to protein functionality has
been data-driven, using machine learning methods trained on sequence data both
predictively and generatively\cite{Russ:2020,Hawkins:2021}. Prominent among the
predictive machine learning models is the AlphaFold\cite{Jumper:2021} project of
DeepMind, a subsidiary of Alphabet Inc. Biophysical models of protein evolution
frequently assume that evolution is determined by the sequences that optimize a
structurally based ``fitness landscape"\cite{Echave:2017,Brooks:2019}. For
example, lattice models of short proteins, for which an exact enumeration of
configurations is feasible, have been used to study the connection between
sequence and specific targeted folded
structures\cite{Dill:1993/99,Shakhnovich:1993,Jacquin:2016}. In the lattice
models, where the dynamics of configurations itself is not well-defined, a
fitness landscape based on a target structure for a sequence is optimized via
random mutations in the sequence space. Rather than being based on dynamical
information, the fitness criterion is determined by the free energy of a Potts
model\cite{Wu:1982}, defined in terms of the adjacency matrix of contacts in the
protein\cite{Wolynes:2014,Jacquin:2016}. In contrast, while the microscopic
dynamics in off-lattice models is clear, the definition of a fitness criterion
for optimal folding is both conceptually and computationally challenging. Part
of the difficulty in investigating the molecular features that determine protein
structure and its connection to dynamics arises from the intractability of
determining how the free energy landscape and dynamical folding pathways depend
on sequence structure and external conditions such as the temperature. Unlike
most off-lattice coarse grained models of biomolecular systems, the structure
and dynamics of discontinuous potential models based on distance constraints can
be examined at any temperature for any choice of interaction energy once the
entropy of the system's states and the distribution of bond distances within
each state are known.

The purpose of this paper is twofold: First, we introduce an efficient
computational approach to evaluate both the configurational entropy and the mean
first passage times for discontinuous potential models, based on adaptive
event-driven sampling. We then present a variational optimization procedure in
the context of a Markov state model to determine the state energies that
minimize the first passage time, subject to a set of structural constraints. In
this case, the first passage time is evaluated for a process in which an initial
state with no bonds evolves to the fully bonded ``native" state under a set of
constraints determined by the thermodynamic requirement of a predominant native
state population.

The outline for this paper is as follows: In Sec.~\ref{sec:cgm}, the
protein-like model is
introduced\cite{Zhou:1997,Zhou2:1997,Zhou:1999,Bayat:2012/245103}. In
Sec.~\ref{sec:sconfig}, the configurational entropy is defined and related to
the thermodynamic structure of the discontinuous potential model. In
Sec.~\ref{sec:msm}, a Markov state model for the simplified dynamics of the
evolution of state populations is introduced. The explicit expressions are given
for elements of the rate matrix that can be computed using only
temperature-independent geometric information. Subsequently, we outline an
adaptive procedure based on event-driven dynamics to evaluate the
configurational entropies, as well as the first passage times, that parameterize
the rate matrix in the Markov state model. Adaptations to the method are
discussed in the two sections that follow. We aim to apply the sampling approach
in a massively parallel framework, and introduce techniques to improve the rate
of convergence in calculations involving states that differ substantially in
their configurational entropy, for which the first passage times are large.
This is followed by Sec.~\ref{sec:equildyn}, with the introduction of a
variational principle to optimize the interactions that lead to rapid folding in
the Markov state model, as well as several key-related measures that are useful
to characterize the folding mechanism. The variational optimization of the
folding time of two different model protein systems---a model of the crambin
protein which is rich in secondary structures, and a small model system that
possesses a native state with a highly-strained helical structure and frustrated
intermediates---is discussed in Secs.~\ref{sec:crambin} and
\ref{sec:frustratedModel}. Finally, concluding remarks are contained in
Sec.~{\ref{sec:summary}.

\section{The coarse-grained model} \label{sec:cgm}

The model we consider here, similar to one introduced by Zhou and
Karplus\cite{Zhou:1997,Zhou2:1997,Zhou:1999}, is based on a coarse-grained
approach in which each amino acid residue of a linear, protein-like chain is
represented by a bead. The chain is immersed in a fluid in thermal equilibrium
at a temperature $T$. We assume that the effect of the fluid is to alter the
energy of the configurations of the chain, and provide a stochastic environment
for the motion of the beads diffusing in the fluid. In the chain, there are
local and nonlocal bonds that connect the beads. Local bonds occur between
nearest and next-nearest neighboring beads. These bonds can correspond to
peptide bonds in the primary structure of a protein. Local bonds are modeled
using an infinite square well potential:
\begin{equation}
    U\left(r_{ij}\right) =
    \begin{cases}
        0 & \text{if}\ \sigma_1 < r_{ij} < \sigma_2 \text{, and} \\
        \infty & \text{otherwise,}
    \end{cases}
\end{equation}
where $U\left(r_{ij}\right)$ is the potential energy of the local bond, $r_{ij}$
is the distance between two nearest or next-nearest neighboring beads $i$ and
$j$, and $\sigma_1$ and $\sigma_2$ are the minimum and maximum bonding
distances, respectively. For nearest neighbors, $\sigma_1 = 1$, which is taken
as the unit of length in the model, and $\sigma_2 =
1.17$\cite{Schofield:2014/095101}. For next-nearest neighbors, $\sigma_1 = 1.4$
and $\sigma_2 = 1.67$ are chosen to restrict the bond angles to be between
$75^{\circ}$ and $112^{\circ}$ to mimic the space that the side chains in amino
acids would normally occupy in a protein.

Non-local bonds occur between beads that are not nearest or next-nearest
neighbors. These bonds account for interactions between the side chains of amino
acids in a protein to form its secondary structures. A nonlocal bond $k$,
formed at a distance $r_{ck}$ between beads $i$ and $j$, is modeled using a step
potential:
\begin{equation} \label{eq:attractivestep}
    U_{k} \left(r_{ij} | \alpha \right) =
    \begin{cases}
        \infty & \text{if}\ r_{ij} < r_h \\
        \epsilon_k(\alpha) & \text{if}\ r_h \leq r_{ij} \leq r_{ck} \\
        0 & \text{if}\ r_{ij} > r_{ck},
    \end{cases}
\end{equation}
where $r_{ij}$ is the distance between the beads. The energy of the bond
$\epsilon_k(\alpha)$ may depend conditionally on the overall configuration
$\alpha$ of the system (i.e., the other nonlocal bonding distances). With this
flexible design of the bonding energy, the model can describe systems with
nonlocal interactions that effectively allow for non-pairwise interactions in
which the energy of a bond depends on the specific configuration involved. In
this way, side chain interactions and temperature-dependent solvent effects,
such as hydrophobicity, can be incorporated into the model in a mean-field way.
At a distance of $r_{ij} = r_h = 1.25$, a hard-core repulsion accounts for
excluded volume interactions. At a distance of $r_{ck}$, a bond forms between
two nonlocal beads, which contributes a factor of $\epsilon_k(\alpha)$ to the
total energy. Note that such a bond is either ``on" or ``off," depending on the
geometric distance between the beads forming the nonlocal bond. If two beads do
not form a nonlocal bond, they will collide elastically at the hard-core
repulsion distance $r_h$.

\section{The thermodynamics and dynamics of the coarse-grained model}
\label{sec:thermo}

\subsection{The configurational entropy} \label{sec:sconfig}

A configuration of a system with $N$ monomer beads is specified by the
$3N$-dimensional vector of bead positions $\vect{R} = (\vect{r}_1, \dots,
\vect{r}_N)$, where $\vect{r}_i$ is the position vector of bead $i$ in the
system. In a model with a step potential and infinite hard wall interactions,
physically allowed configurations $\vect{R}$ of the system must satisfy distance
constraints that force nearest and next-nearest beads in the chain to be within
a short distance of one another determined by $\sigma_1$ and $\sigma_2$. The
entire configurational space of allowed configurations is geometrically
partitioned into states of the system by the set of $n_b$ nonlocal bonding
distances $\{r_{c_k} | k = 1, \dots, n_b\}$. A configurational state $c$ can be
represented as a binary string:
\begin{equation} \label{eq:config}
    c = c_1 \dots c_{n_b},
\end{equation}
where each term $c_i$ in the string $c$ takes on a binary value of $1$ if $x_i <
r_{ci}$ and $0$ if it is not. For example, for a model with three nonlocal
bonds, the configuration $000$ refers to an unfolded chain with no nonlocal
bonds.

To develop the statistical mechanics of the model, we define the indicator
function for a configurational state $c$,
\begin{equation}
    \label{eq:indicator}
    \mathbbm{1}_c\left(\vect{R}\right) =
    \begin{cases}
        1 & \text{if all constraints for $c$ are satisfied, and} \\
        0 & \text{otherwise.}
    \end{cases}
\end{equation}
The partitioning of the $3N$-dimensional space of microscopic configurations
enables us to reduce the large number of allowed configurations (equal to the
volume of the configurational space) to a finite and discrete set of $n_s =
2^{n_b}$ coarse-grained states for which structural and dynamical properties can
be derived.

The coarse-grained model is unusual in that the dimensionless nonideal entropy,
$S_c$, defined by
\begin{equation} \label{eq:entropy}
    S_c = \ln \left(\frac{1}{V^N} \int \mathbbm{1}_c \left(\vect{R}\right)
    d\vect{R} \right),
\end{equation}
can be determined entirely by the distance constraints between the beads for any
configuration $c$. The integral of $\mathbbm{1}_c$ over the volume of
configurations can be viewed as the volume of the subspace occupied by
configuration $c$ in the full configurational space.

For a given model with a prescribed set of interaction energies $\{\vect{E}\}$,
the canonical probability, $P_c$, of a configuration $c$ with a potential energy
$E_c$ at an inverse temperature $\beta^*$ is
\begin{equation} \label{eq:prob}
    P_c = \langle \mathbbm{1}_c \rangle = \frac{e^{-\beta^* E_c}
    e^{S_c}}{\sum_{\alpha=1}^{n_s} e^{-\beta^* E_{\alpha}} e^{S_{\alpha}}}
    = \frac{e^{-\beta^* F_c}}{\sum_{\alpha=1}^{n_s} e^{-\beta^* F_{\alpha}}}.
\end{equation}
Here, $\langle \cdots \rangle$ denotes the canonical ensemble average, $n_s =
2^{n_b}$ is the total number of configurations, and $F_c = E_c - T^*S_c$ is the
free energy of configuration $c$.

As evident in Eq.~\eqref{eq:entropy}, the entropy difference between two states
is independent of both the temperature of the system and the set of interaction
energies. As a result, once the configurational entropy for all states has been
determined, the canonical probability of any state for {\it any} choice of
interaction energies at any temperature can be evaluated. This flexibility
permits us to examine not only how the morphology of the free energy landscape
changes with temperature, but also how changing the interaction energies of
different states, which is similar to changing the molecular identity of each
bead, influences the thermodynamics of the system. The generality of the model
also allows for the effects of different interaction energies on aspects of the
dynamics, such as the structural folding time, to be examined.

\subsection{The transition rate matrix and the mean first passage times}
\label{sec:msm}

In a viscous fluid environment, there is a separation of time scale between the
typical time for a change of configuration of a protein and the time it takes to
equilibrate locally in each state. Under such conditions, the evolution of
populations of configurations at intermediate time scales that are long compared
to the molecular time scale, but much shorter than the overall folding time, can
be described by a Markov state model\cite{Schofield:2014/095101}. The dynamics
can describe the folding process as a series of transitions between
configurations, defined in Eq.~\eqref{eq:indicator}, that differ by one bond;
such transitions represent a structural change in the protein as bonds form or
break.

In a Markov state model, a population of configurations, $\vect{P}(t) =
\{P_1(t), \dots, P_{n_s}(t)\}$, evolves according to the continuous time
Markovian dynamics,
\begin{equation}
    \frac{d\vect{P}(t)}{dt} = \vect{K} \cdot \vect{P}(t),
    \label{eq:markovDynamics}
\end{equation}
where $\vect{K}$ is the transition rate matrix. The off-diagonal elements of
$\vect{K}$ are the time-independent rates of transitioning from one state to
another. Consider the case where states are ordered in index from fewest bonds
to most bonds, and suppose $j > i$ is formed from $i$ by the addition of a
single bond. Then, due to diffusive barrier crossing, the inverse of $K_{ji}$ is
of the form\cite{Schofield:2014/095101}
\begin{align} \label{eq:invKji}
    K_{ji}^{-1} &= e^{-\beta^* F_{ij}} \tau^{-}_{(ij)} + \tau^{+}_{(ij)} \\
    &= \frac{P_i}{P_j} \tau^{-}_{(ij)} + \tau^{+}_{(ij)}, \nonumber
\end{align}
where $\tau^{-}_{(ij)}$ and $\tau^{+}_{(ij)}$ are the mean inner and outer
equilibrium first passage times for the pair of states $i$ and $j$: That is,
$\tau^{+}_{(ij)}$ corresponds to the time required for a pair of beads, whose
initial separation $r_{ij}$ is greater than the transition state value $r_c$, to
diffuse to $r_c$, averaged over a (conditional) equilibrium distribution of
initial separations. Correspondingly, $\tau^{-}_{(ij)}$ is the mean first
passage time to $r_c$ for beads averaged over an equilibrium distribution of
initial distances $r_{ij} < r_c$. When the dynamics of beads in the solvent is
diffusive, the mean first passage times for the transition from $i$ to $j$ can
be estimated as\cite{Schofield:2014/095101}
\begin{align} \label{eq:mfpt+}
    \tau^{+}_{(ij)} &= \frac{1}{D_{(ij)} }\int_{r_c}^{r_{\text{max}}} \frac{(1 -
    C^{+}_{(ij)}(r))^2}{\rho^{+}_{(ij)}(r)} dr \\
    \tau^{-}_{(ij)} &= \frac{1}{D_{(ij)}} \int_{r_{\text{min}}}^{r_c} \frac{
    C^{-}_{(ij)}(r)^2}{\rho^{-}_{(ij)}(r)} dr,
    \label{eq:mfpt-}
\end{align}
where $D_{(ij)}$ is the self-diffusion coefficient for the relative distance $r$
between beads involved in the bond that is formed or broken between states $i$
and $j$ in the solvent, $\rho$ is the probability density of the bonding
distance, and
\begin{align}
    C_{(ij)}^{-}(r) &= \int_{r_{\text{min}}}^r \rho^{-}_{(ij)}(x) \, dx
    \nonumber \\
    C_{(ij)}^{+}(r) &= \int_{r_c}^{r} \,\rho^{+}_{(ij)}(x) \, dx \nonumber
\end{align}
are the respective cumulative distributions of the distances $r$. The constants
of integration $r_{\text{min}}$ and $r_{\text{max}}$ correspond to the minimum
and maximum distances that can separate a pair of nonlocally bonding beads, and
they can be taken to be zero and infinity, respectively, since the integrand
vanishes in both limits. Generally speaking, the self-diffusion coefficients
$D_{(ij)}$ depend on both the solvent friction as well as the internal friction
that arises from the particular distance constraints determining the states $i$
and $j$. By construction, the Markov state model obeys detailed balance,
\begin{equation} \label{eq:detailedbal}
    K_{ij} P_j = K_{ji} P_i,
\end{equation}
and Eq.~\eqref{eq:markovDynamics} has a unique stationary equilibrium
distribution of populations.

\begin{figure}[htbp]
    \centering
    \includegraphics[height=5.5cm]{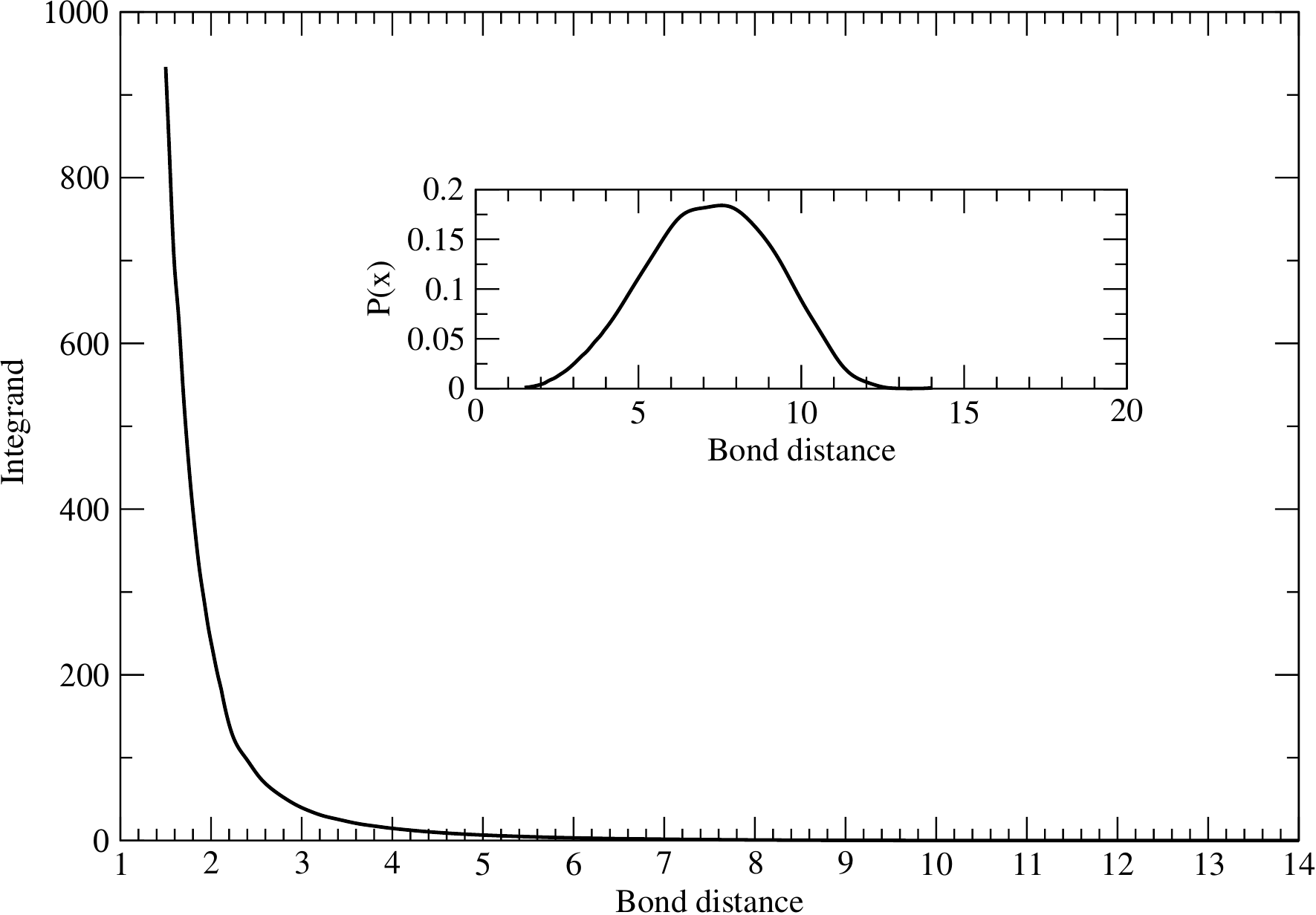}
    \caption{The integrand of the outer mean first passage time $\tau^+$ for a
    40-bead system for a nonlocal bond between beads separated by $20$
monomers. The inset shows the probability density $\rho^{+}(r)$ of the bonding
distance.}
    \label{fig:staircaseFPT}
\end{figure}

Note that the relative probability of states $i$ and $j$,
\begin{equation}
    \frac{P_i}{P_j} = \frac{e^{-\beta^* E_i} e^{S_i}}{e^{-\beta^* E_j} e^{S_j}}
    = e^{-\beta^* \left(F_i - F_j\right)}, \label{eq:relativeProbability}
\end{equation}
plays an important role in determining the transition rates, but it does not
affect the mean first passage times. In the low temperature limit, $P_j \gg
P_i$, since state $j$ has an additional bond relative to state $i$ and hence
$K_{ji} \approx 1/\tau_{(ij)}^+$. Under these conditions, the rate of back
transitions, $K_{ij}$, will be exponentially small since
\begin{eqnarray*}
   K_{ij} &=& \frac{P_i}{P_j} K_{ji} \approx e^{-\beta^* F_{ij}}
   \frac{1}{\tau^{+}_{(ij)}}.
\end{eqnarray*}

The adaptive algorithms that we present in the following sections generate bond
distances distributed according to the conditional equilibrium densities
$\rho^{+}(r)$ and $\rho^{-}(r)$ during the iterative process. Using the set of
recorded distances, the smooth fit of the probability density $\rho(r)$ and the
cumulative distribution functions $C(r)$ are constructed in one of two ways:
Either the empirical cumulative distribution is expanded in an orthonormal
basis\cite{VanZonSchofield:2010}, or alternatively, a maximum-likelihood
estimate of an expansion of the logarithm of the density is constructed using
splines\cite{Schofield:2017}. Both approaches make use of goodness-of-fit
statistical tests to judge the quality of the fit. In
Fig.~\ref{fig:staircaseFPT}, the integrand for the outer first passage time for
the transition between states in a $40$-bead system is shown as a function of
the bonding distance, as well as the probability density $\rho^{+}(r)$. The
latter is constructed from a continuous spline fit of the logarithm of the
density, or the potential of mean force. With these continuous and smooth
functions in hand, the mean first passage times in Eq.~\eqref{eq:mfpt+}
and~\eqref{eq:mfpt-} and the elements of the transition rate matrix $\vect{K}$
are easily evaluated numerically using Gaussian quadrature. Note that the
integrand of the outer first passage time is determined primarily by the fit of
the density in the transition region near $r_c$ where the integrand is largest.

\section{Numerical solution of the configurational entropy and first passage
time} \label{sec:numsol}

In a complex system with a large number of beads, the shapes of the sub-volumes
of different configurations in the high dimensional space are complicated,
making Eq.~\eqref{eq:entropy} impossible to evaluate exactly for all but the
simplest models. To compute the configurational entropies for larger model
chains, we must resort to using Monte Carlo (MC) methods. In the next several
sections, the algorithm used to compute the configurational entropy and first
passage times is detailed. A brief overview of the procedure is as follows:
\begin{enumerate}
        \item An ensemble of initial structures with no bonds is generated by
            selecting a set of bond distances, bond angles, and dihedral angles
            from the appropriate distribution of values. Configurations
            violating any distance constraints are rejected. The distribution of
            the distances of nonlocal interactions are used to estimate the
            first passage time to form each of the possible nonlocal bonds for
            the set of structures with a single bond (see Eq.~\eqref{eq:mfpt+}).
        \item The set of states defined by which nonlocal bonds are active is
            partitioned into layers based on the number of nonlocal bonds. The
            layer method is detailed in Sec.~\ref{sec:layer}.
        \item For states for which the estimate of the mean first passage time
            from earlier calculations is large ($\tau^+ > 10$), the attractive
            step potential depending on the active bond distance is replaced by
            a staircase potential in Sec.~\ref{sec:staircase} to reduce the
            computational cost.
        \item The difference in entropy and the first passage times between
            pairs of states in adjacent layers that differ by only one bond are
            computed in parallel using an adaptive Monte Carlo sampling
            algorithm (see Sec.~\ref{sec:mc}) combined with replica-exchange
            between simulations at fixed intervals.
        \item In the adaptive procedure, rejection-free Monte Carlo updates are
            carried out using event-driven dynamics in a hybrid Monte Carlo
            algorithm.
            \begin{enumerate}
                \item The initial phase of the evaluation of the entropy
                    difference consists of a fixed number ($10^7$) of adaptive
                    adjustments of the entropy using a Wang--Landau algorithm
                    (Eq.~\eqref{eq:wl1}).
                \item Convergence of the entropy difference is assessed by
                    applying the event-driven dynamics with fixed entropy values
                    to generate a set of (independent) configurations and
                    applying the G-test for uniformity given in
                    Eq.~\eqref{eq:Gtest}.
                \item If the G-test is not satisfied, the entropy values are
                    adjusted using Eq.~\eqref{eq:biasedentropy}, and the
                    previous step is repeated until the test is satisfied.
            \end{enumerate}
        \item The distances between beads forming nonlocal bonds, which are
            recorded at regular time intervals throughout the simulation, are
            used to obtain the mean first passage times given in
            Eq.~\eqref{eq:mfpt+} and~\eqref{eq:mfpt-}.
        \item The biased entropies and mean first passage times can then be used
            to construct the transition rate matrix in the Markov state model
            for a choice of interaction energies (see Eq.~\eqref{eq:invKji}).
\end{enumerate}

\subsection{Adaptive Monte Carlo sampling} \label{sec:mc}

For a molecular system suspended in a solvent in thermal equilibrium at inverse
temperature $\beta$, the configurations of the molecule are canonically
distributed. For a given model with a set of $n_s$ energies $\{\beta^*
\vect{E}\}$, the probability of a configuration in the ensemble is given by
Eq.~\eqref{eq:prob}. The entropy difference between states $i$ and $j$ obeys
\begin{equation}
    e^{S_i - S_j} = e^{S_{ij}} = \frac{e^{\beta^* E_i} P_i}{e^{\beta^* E_j} P_j},
    \nonumber
\end{equation}
and, hence,
\begin{equation}
    S_{ij} = \ln \left(\frac{P_i}{P_j}\right) + \beta^* \left(E_i - E_j\right).
    \nonumber
\end{equation}
Suppose $n_c$ samples of states are drawn {\it independently} with a canonical
probability for a model with a set of energies $\{\beta^* \vect{E}\}$. The
number of states of type $i$ in the sample is denoted by $n_i$. Using the
empirical probability of state $i$, $\hat{P}_i = n_i/n_c$, an estimator of the
entropy difference between states $i$ and $j$ can be defined as
\begin{equation} \label{eq:estimator}
    \hat{S}_{ij} = \ln \left(\frac{n_i}{n_j}\right) + \beta^*
    \left(E_i - E_j\right).
\end{equation}
If the states $\{1, 2, \dots, s\}$ are sampled independently, the set of counts
$\{n_1, n_2, \dots, n_s\}$ is multinomially distributed with probability
\begin{equation}
    P \left(\{n_1, n_2, \dots, n_s\}\right) = \frac{n_c!}{n_1! \dots n_s!}
    P_1^{n_1} P_2^{n_2} \dots P_s^{n_s}. \nonumber
\end{equation}
The mean and the variance of the entropy estimator in Eq.~\eqref{eq:estimator}
are
\begin{align} \label{eq:sijavg}
    \langle \hat{S}_{ij}\rangle &=
    S_{ij} + \frac{1}{2 n_c} \left(\frac{1}{P_j} - \frac{1}{P_i}\right) + O
    \left(\frac{1}{n_c^2}\right) \\
    \sigma_{S}^2 &= \frac{1}{n_c} \left(\frac{1}{P_i} + \frac{1}{P_j}\right) +
    O\left(\frac{1}{n_c^{2}} \right) \geq \frac{2 n_s}{n_c},
\end{align}
where all $P_k > 0$, which implies that $\langle \hat{S}_{ij} \rangle$ converges
to $S_{ij}$ as $n_c^{-1} \rightarrow \infty$. Note that the rate of convergence
of the estimator is optimized when $P_i \approx P_j$, at which point the minimum
value of the variance is $2 n_s / n_c$. The minimum variance is achieved when
$\beta^* E_i = S_i$. This choice of the set $\{\beta^* \vect{E}\}$ is not known
{\it a priori} and must be determined self-consistently, as discussed below.
Note that other choices of $\{\beta^* \vect{E}\}$ may result in $P_i \gg P_j$,
in which case the empirical average of the entropy converges slowly due to a
large standard error given that $1 / {(n_c P_j)} \gg 1$.

The estimator Eq.~\eqref{eq:estimator} requires a set of samples drawn from the
canonical ensemble. Metropolis Monte Carlo (MMC) algorithms are an appealing
sampling approach to generate a sample of states since they do not require
computing normalizing factors to generate states with known probabilities.
However, efficient implementations of the MMC algorithm require proposing trial
configurations from the current state that are both statistically likely and yet
differ significantly. For chain molecules, particularly those that have excluded
volume constraints, this is a difficult task\cite{Schofield:1998}, although
methods using crankshaft rotations\cite{Verdier:1962,Kumar:1988},
configurational bias regrowth\cite{Seipmann:1992,Frenkel:1992}, and normalizing
flows\cite{Rezende:2015} exist to generate global changes to configurations.

The principal challenge in efficient MMC sampling in this context is the highly
correlated way in which configurations must change to generate states of high
probability. Dynamical sampling methods that evolve all degrees of freedom
provide a viable solution for the rapid exploration of local structures. Here,
we use the hybrid Monte Carlo (HMC) method to generate configurations with a
canonical probability based on a dynamical updating scheme\cite{Duane:1987}. In
this procedure, the dynamical updates must be time-reversible and must conserve
phase space volume. In most applications, proposed configurations are generated
by numerically solving the equations of motion for a given potential using
symplectic split-operator integration schemes. For systems interacting via
discontinuous step potentials and hard walls, the equations of motion are
exactly solvable (within numerical precision), and the dynamics of the system is
time-reversible and conserves phase space volume.

In our implementation of the HMC scheme, the current configuration $\vect{R}$ is
augmented with momenta $\vect{P}$ drawn from a normal distribution with zero
mean and unit variance so that the system acquires a kinetic energy $K(\vect{P})
= \vect{P}^2/2$. Then, the system is propagated forward for a time interval
$\tau_p$ with Hamiltonian dynamics from an initial state $\vect{X} = (\vect{R},
\vect{P})$ to a final state $\vect{X}_{\tau_p} = (\vect{R}_{\tau_p},
\vect{P}_{\tau_p})$. The Hamiltonian $H(\vect{R}, \vect{P})$ is the sum of the
kinetic energy $K(\vect{P})$ and a discontinuous potential $U(\vect{R})$. The
final configuration $\vect{R}_{\tau_p}$ of the trajectory is then accepted or
rejected as the next state in a Markov chain with acceptance probability given
by
\begin{equation}
    A(\vect{X} \rightarrow \vect{X}_{\tau_p}) = \text{min} \left(1, e^{-\Delta
    H}\right),
\end{equation}
where $\Delta H$ is the difference between the final and initial Hamiltonians.
When event-driven dynamics generate trial configurations $\vect{R}_{\tau_p}$,
the HMC algorithm proposes updates in a rejection-free manner since the
Hamiltonian is exactly conserved so that $\Delta H = 0$, and the probability of
acceptance of a trial configuration is unity. For discontinuous potential
systems, the dynamic sampling trajectories are solved exactly (i.e., within
numerical precision) using event-driven simulation methods, and efficient
implementation of event-driven dynamics should make use of event trees, hybrid
queues, and other cost-saving techniques\cite{Rapaport:2004}. The sampling
procedure generates a set of states $\vect{R}$ asymptotically distributed with
probability proportional to $e^{-U(\vect{R})}$. Instead of using an actual
physical potential $U(\vect{R})$ to govern the dynamical updates in the Monte
Carlo procedure, we use an estimate of the entropy $U(\vect{R}) = S_b(\vect{R})$
that approximates the true entropy $S(\fakebold{\mathbbm{1}} (\vect{R}))$,
where $\mathbbm{1}_c(\vect{R})$ is the indicator function for state $c$ defined
in Eq.~\eqref{eq:indicator}. The HMC sampling procedure generates a Markov chain
of states in which configuration $i$ appears with probability
\begin{equation}
    P_i(\vect{S}_b) = \frac{e^{-S_{i,b}} e^{S_i}}
    {\displaystyle{\sum_{k=1}^{n_s} e^{-S_{k,b}} e^{S_k}}} \approx
    \frac{1}{n_s}. \label{eq:hmc}
\end{equation}

In order to ensure that the $n_c$ samples are drawn independently, the time
$\tau_s$ between recording configurations of the system should be larger than
the largest outer mean first passage time between states when only local,
dynamic updates are used to propose trial configurations. In this case, $\tau_s$
is set to be a multiple of the basic short propagation time $\tau_p$ of the
dynamical updates.

However, in Eq.~\eqref{eq:hmc}, the optimal values of the set of the biasing
potential $\vect{S}^*_{b} = \{S^*_{i,b} | i = 1, \dots, n)\}$ that lead to a
uniform sampling are not known {\it a priori} and must be determined iteratively
using an adaptive procedure. A number of adaptive methods that are effectively
equivalent have been proposed in the literature to address this problem,
including the Wang--Landau algorithm in its many flavors\cite{WangLandau:2001},
well-tempered metadynamics\cite{Laio:2002,Barducci:2008}, and
self-healing umbrella sampling\cite{Marsili:2006}.

The essential idea of the adaptive procedure is to construct a sequence of
configurations $\{{\bf{R}}_i | \vect{S}_b^{(n)}\}$, in which each of the states
${\bf{R}}_{n+1}$ is obtained from the previous state ${\bf{R}}_n$ by applying an
evolving transition matrix $\vect{T}(\vect{S}_b^{(n)})$. The parameters
$\vect{S}_b^{(n)}$ are determined by a difference equation of the form
\begin{align}
    \vect{S}_b^{(n+1)} &= \vect{S}_{b}^{(n)} + \gamma_{n+1} \,
    \vect{f}({\bf{R}}_{n+1} | \vect{S}_b^{(n)}) \label{eq:wl1} \\
    &= \vect{S}_{b}^{(n)} + \gamma_{n+1} \, \vect{h}(\vect{S}_b^{(n)}) +
    \gamma_{n+1} \, \vect{H} ({\bf{R}}_{n+1} | \vect{S}_b^{(n)}), \nonumber
\end{align}
where $\gamma_{n+1}$ is a decreasing function of $n$ and the adaptive function
$\vect{f}$ penalizes visits to the current state and encourages visits to other
states. In Eq.~\eqref{eq:wl1}, $\vect{h}$ is the mean drift in the difference
equation at index $n$, and $H$ is the fluctuation around the mean.

The various algorithms differ in their choice of both the dependence of
$\gamma_n$ on the number of steps $n$, and the form of the adaptive function
$\vect{f}$. Here, we use the commonly chosen adaptive function $f(\vect{R}) =
\fakebold{\mathbbm{1}} (\vect{R})$ that penalizes future visits to the current
state by increasing its entropy by $\gamma_{n+1}$. In general, the convergence
of the sequence of biases $\{\vect{S}_b^{(n)}\}$ to a unique fixed-point
solution $\vect{S}_b^{*}$ is difficult to establish for a particular choice of
$\gamma_n$, $\vect{f}$, and the transition matrices $\vect{T}$, but it has been
proved for the algebraic protocol $\gamma_n = \gamma^* / n^\alpha$, where
$\alpha \in (1/2, 1]$, provided the transition matrices are sufficiently
mixing\cite{Fort:2012,Fort:2017}. For example, for the parameter choice
$\gamma^* = n_s$ and $\alpha = 1$, it has been shown\cite{Fort:2012} that the
sequence $\{\vect{S}_n^{(n)}\}$ converges to $\vect{S}_b^*$ as $n^{-1}$, and
that the set $\vect{S}^{(n)}_b$ has a multivariate normal distribution with a
mean $\vect{S}^*_b$ and a covariance matrix proportional to $\vect{U}_t = n_s
\gamma_n \vect{U}^* = n_s/t \, \vect{U}^*$, where $t = n/n_s$ is the state
size-dependent scaled time between updates of $\gamma_n$. Here, $\vect{U}^*$ is
a covariance matrix that depends on the fluctuations of $H$ determined by the
sequence of transition matrices $\vect{T}$. As a result, it is difficult to
estimate $\vect{U}^*$ to determine the standard errors of the entropy values
$\vect{S}_b$ in the adaptive procedure.

To assess the accuracy of the configurational entropy, we iterate
Eq.~\eqref{eq:wl1} for a fixed number of total updates $t_f = m t$ with a large
value of $m = 10^7$. At this point, a series of $n_c$ independent trajectories
are generated using the final set of biases and the number $n_i$ of counts of
uncorrelated states $i$ recorded. The recorded empirical distribution of states
is then checked for uniformity using a statistical test. Here, we use the G-test
based on the statistic
\begin{equation}
    G = -\frac{2}{q_2} \sum_{i=1}^{n_s} n_i \, \ln \left(\frac{n_i}{e_i}\right),
    \label{eq:Gtest}
\end{equation}
where $e_i = P_i n_c = n_c/n_s$ is the expected number of counts of state $i$
when $P_i$ is uniform and the term\cite{Smith:1981}
\[
    q_2 = 1 + \frac{n_s+1}{6n_c} + \frac{n_s^2}{6n_c^2}
\]
corrects for small sample sizes. When the sample counts $\{n_i\}$ are
independent, the G-statistic is asymptotically $\chi^2$-distributed with $n_s -
1$ degrees of freedom, allowing the p-value of the computed statistic to be
evaluated.

Convergence has been achieved when $p > p_c$ and the distribution of
configurations is considered statistically consistent with a uniform
distribution. If $p < p_c$, the biased entropy values in iteration $n$ can be
updated according to
\begin{equation} \label{eq:biasedentropy}
    S_{i, \text{b}}^{(n+1)} = S_{i, \text{b}}^{(n)} + \ln \left(
    \frac{n_i}{n_{c}}\right).
\end{equation}
Strictly speaking, this additional iterative process is not required since the
estimator for the configurational entropy is unbiased and has a variance that is
close to optimal since the biased probabilities are already close to uniform,
$P_i \approx P_j$. If desired, another iteration of the sampling can be
performed with the updated bias values until convergence is obtained and $S_i =
S_{i, \text{b}}$ within statistical resolution. More stringent statistical tests
for convergence can be applied if desired. For example, after the process has
passed the condition $p > p_c$, the actual distribution of a set of G-statistics
from independent runs can be tested against a $\chi^2$-distribution using a
goodness-of-fit test, such as the Kolmogorov--Smirnov test\cite{Marsaglia:2003}.
It should be emphasized that the failure of the condition $p > p_c$ does not
necessarily indicate that the data of counts are inconsistent with a multinomial
distribution, since the statistical test also relies on the assumption that the
samples are drawn independently.

The final configurational entropy difference between states $i$ and $j$ with
counts $n_i$ and $n_j$ is
\[
    \Delta S_{ij} = \Delta S_{ij, b}^{(n)} + \ln(n_i/n_j) \pm \sqrt{\frac{A}{n_i
    + n_j}},
\]
where samples are recorded at time intervals $\tau > n_b^2\tau^+$ and $A$ is the
upper percentile value of the $\chi^2$-distribution with $n_s - 1$ degrees of
freedom\cite{Goodman:1965}. The length of the production run required for a
given statistical resolution can be estimated using confidence intervals for
multinomial proportions\cite{Goodman:1965,May:2000}.

During each iteration, the bonding distances between all beads for each state
$i$ of the $n_s$ states explored can be used to calculate the mean first passage
times for state $i$. The bonding distances are distributed with the conditional
equilibrium density for this state due to the fact that the bias
$S_{i,\text{b}}$ is the same for all configurations in state $i$.

It is important to emphasize that the statistical analysis presented above
assumes that each sampled configuration is drawn independently from the
canonical probability density with a corresponding multinomial distribution of
states. In practice, this will not be the case when local Monte Carlo proposals
alone are used, since the proposed trial configurations are highly correlated
with the current state. Correlations exist when the lengths of the trajectories
$\tau_s$ are not long enough to generate independent configurations in the
Markov chain. If the trajectory segments are only long enough to form or break a
single bond, the overall dynamics in the state space is diffusive at best and
the statistical tests for uniformity are inappropriate. Under these
circumstances, the correlation time of the state counts $\{n_i\}$ must be
analyzed to insure that successive states used in the convergence test are
independent.

\subsection{The layer simulation method} \label{sec:layer}

If the protein being modeled can form many bonds $n_b$, the number of possible
states and the number of configurational entropy values $n_s = 2^{n_b}$ to be
computed will be large. As $n_s$ increases, the covariance matrix $\vect{U}_t$
of the adaptive procedure, which scales quadratically with $n_s$, becomes large.
The convergence of the entropy to $\vect{S}_b^*$ will therefore be very slow,
particularly when some of the transitions are infrequent due to long first
passage times. The states generated via short trajectories remain correlated for
increasingly long periods of time as the number of bonds increases. For example,
when the full set of $n_s$ states are sampled using local dynamical updates, the
dynamics of the state space in the limit where the states are generated with
uniform probability obeys a Master equation of the form
\begin{eqnarray*}
    \frac{d P_i(t)}{dt} = \kappa \left(P_{i+1}(t) + P_{i-1}(t) - 2 P_i(t)
    \right),
\end{eqnarray*}
where $\kappa = 1 / (\tau^{-} + \tau^{+}) \sim 1 / \tau^{+}$ is the rate of
transitions to neighboring states. These dynamics generate a uniform
distribution of states $P_i \sim 1 / n_s$ on time scales governed by the
relaxation modes $\lambda_m = 2\kappa \sin^2 \big(m\pi / (2(n_b+1))\big)$.
States remain correlated for time scales up to the overall equilibration time
$\tau_{eq} \sim 1/\lambda_1 \sim (8/\pi) n_b^2 \tau^+$, where $n_b$ is the total
number of bonds that can be formed. Thus, to generate uncorrelated samples
uniformly, the length of trajectories $\tau_s = s \tau_p$ should be scaled by
$n_b^2$, relative to two-state models for which trajectories of length $\tau^+$
are adequate.

To improve the rate of convergence of $\vect{S}_b^{(n)}$ to $\vect{S}_b$, we
consider a layered simulation approach, in which short calculations are
conducted in parallel to sample two states at a time differing by a single
bond. If we define the layer $\ell$ to be the ${n_b} \choose {\ell}$ states in
which there are $\ell$ nonlocal bonds that have formed and $n_b - \ell$ bonds
that have not, each state in layer $\ell$ can lead to $n_b - \ell$ states in
layer $\ell + 1$ by the formation of a single new bond. The pairing of all
states connected in adjacent layers leads to a total set of $n_b n_s/2$ pairs of
connected states for which the difference in entropy is computed. If the entropy
of a configuration is defined relative to the non-bonded state in layer $0$, the
entropy $\Delta S(\alpha_0, \alpha_\ell)$ of a particular state $\alpha_\ell$ in
layer $\ell$ can be estimated by the sum of the entropy differences between
states in adjacent layers in a path that connects state $\alpha_0$ to the state
$\alpha_\ell$:
\begin{align*}
    \Delta S(\alpha_0, \alpha_\ell | \{\alpha_i\}) &= \Delta S(\alpha_0,
    \alpha_1) + \Delta S(\alpha_1, \alpha_2) + \dots \\
    &+ \Delta S(\alpha_{\ell-1}, \alpha_\ell),
\end{align*}
where the path $\{\alpha_i\}$ used is $\alpha_0 \rightarrow \alpha_1
\rightarrow \alpha_2 \rightarrow \cdots \rightarrow \alpha_{\ell-1} \rightarrow
\alpha_\ell$. However, when all states are dynamically connected and none are
geometrically prohibited, there are a total of $\ell!$ unique paths that
connect $\alpha_0$ and $\alpha_\ell$, so a more precise estimate can be obtained
by averaging over all paths that connect the same initial and final states,
\begin{equation}
    \Delta S(\alpha_0, \alpha_\ell) = \frac{1}{\ell!} \sum_{\{\alpha_i\}}
    \Delta S(\alpha_0, \alpha_\ell | \{\alpha_i\}). \label{eq:layerEstimator}
\end{equation}
When each of the computations of $\Delta S(\alpha_i, \alpha_{i+1})$ has
converged and the probability of the states $\alpha_i$ and $\alpha_{i+1}$ is
the same, the mean of the estimator defined in Eq.~\eqref{eq:estimator} is zero
with variance $4/n_c$, where $n_c$ is the number of event-driven trajectories
used to sample the states in the simulation. Hence, the variance of the
estimator for a state in layer $\ell$, Eq.~\eqref{eq:layerEstimator}, is
\begin{equation}
    \sigma^2_{\Delta S_\ell} \geq \frac{4}{(\ell-1)! \, n_c}.
    \label{varianceDeltaS}
\end{equation}

This estimate is useful to determine the number of independent configurations
$n_c$ chosen per iteration for a given level of precision.
\begin{figure}[htbp]
    \centering
    \includegraphics[width=0.75\linewidth]{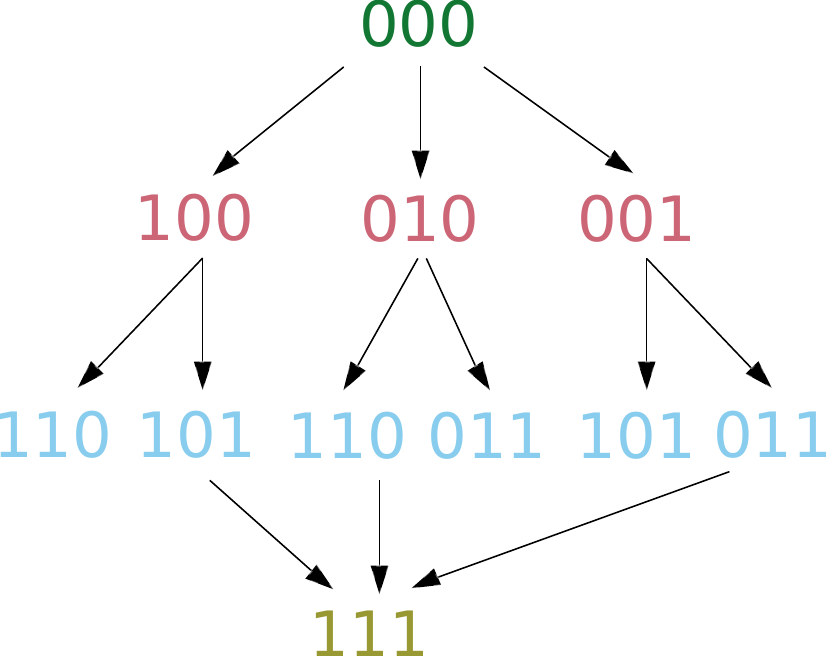}
    \caption{The layer approach for a $20$-bead, three-bond model.}
    \label{fig:layers}
\end{figure}
If the relative precision is set to $5\%$, then the number of sampled states
should be chosen to be larger than $n_c \geq 1600/\Delta S_{m}$, where $\Delta
S_m$ is an estimate of the minimum increase in entropy obtained by breaking one
of the bonds. Typically, for the models considered here, this quantity is
roughly unity (though often it is three times larger).

To illustrate the layer method, consider a three-bond model for which $n_s = 8$.
We present the layer approach in Fig.~\ref{fig:layers}.

The initial state of the chain is linearly extended and entirely devoid of
nonlocal bonds, which is represented by the binary string $000$. The
configuration $000$ makes up layer $0$ (in green). To obtain the configurations
in layer $1$ (in red) with a single nonlocal bond, we run three simulations: one
in which we transition from $000$ to $100$ by forming the first bond, one from
$000$ to $010$ by forming the second bond, and one from $000$ to $001$ where the
third bond is formed. In the event-driven dynamical sampling of the two
designated states, the active nonlocal bond that can be formed or broken is
treated normally with dynamics governed by the step potential (see
Eq.~\eqref{eq:attractivestep}), while the state of all other nonlocal bonds is
fixed by an elastic collision at $r_c$ (i.e., the step potential for these bonds
is infinite and positive). To obtain the configurations in layer $2$ (in blue),
we start with each configuration in layer $1$ and turn on one of each of the two
remaining bonds. Thus, between the first and the second layer we have a total of
${{3} \choose {1}} = 6$ simulations: $100$ to $110$, $100$ to $101$, $010$ to
$110$, $010$ to $011$, $001$ to $101$, and $001$ to $011$. In each of the
simulations, each existing bond in layer $1$ is fixed and is not allowed to
break. Overall, there are a total of $n_b n_s/2 = 12$ computations of
configurational entropy differences. To compute the entropy of a state in layer
$2$ relative to the non-bonded state, the average is taken over the paths
connecting it to state $000$. For example, $\Delta S_{110, 000} = (\Delta
S_{110, 100} + \Delta S_{100, 000} + \Delta S_{110, 010} + \Delta S_{010, 000})
/ 2$.

To avoid quasi-ergodic sampling issues in which transitions between different
types of structures for a given state are rare, the layer simulations are
coupled together by replica-exchange Monte Carlo
moves\cite{Geyer:1991,Geyer:1995,Neal:1996}, in which configurations are
exchanged between adjacent layers with unit probability when they satisfy the
same bonding constraints. For example, a layer simulation connecting a state in
layer $i-1$ with a state in layer $i$ that differs by a single bond can be
coupled to a simulation between a pair of configurations in layers $i$ and $i+1$
that also differ by a single bond. The replica-exchange swaps between the Markov
chains are accepted when both simulations are in states in layer $i$ and
therefore satisfy the same bonding pattern. The swap moves should be attempted
frequently to optimize the efficiency of the replica-exchange
sampling\cite{Opps:2001,Roitberg:2008}. Thus, a given set of configurations are
exchanged with a frequency of 25\% if each layer simulation consists of two
states that only differ by a single bond. The exchange frequency can be
increased by increasing the number of layers that are explored in a given chain.
Similar replica-exchange algorithms have been proposed in the context of the
Wang--Landau algorithm\cite{Vogel:2013,Moreno:2022}. An alternative parallel
implementation\cite{Bornn:2013} of the Wang--Landau algorithm that requires
frequent communication between stochastic trajectories uses an adaptive function
$\vect{h}$. This function depends on a mean number of visits to update a shared
set of biases $\vect{S}_{b}^{(n)}$. In a serial approach, population Monte Carlo
algorithms which generate pools of different structures for a pair of states and
uniformly select a structure from the pool to be updated can accomplish the same
task\cite{Elvira:2017}.

As before, the bonding distances can be recorded and used to compute $2n_b$
inner or outer mean first passage times. The outer first passage time $\tau^+$
between a source and destination state in the next layer can be also used to
estimate the length of a trajectory $\tau_s \sim \tau^{+}$ to generate a
statistically independent configuration in the next layer, where the destination
state is a source state for the next layer. This information is useful in two
ways. First, the computational cost of the procedure can be optimized by
adapting the trajectory length $\tau_s$ to the pair of states. Second, problems
can indicate when a pair of states either are not connected due to geometrical
constraints that are impossible to satisfy, or require unreasonably long
trajectories due to a large value of $\tau^{+}$. In Sec.~(\ref{sec:staircase}),
a biasing procedure is introduced to mitigate the problems associated with large
first passage times.

Another issue that arises for models with a large number of bonds is that some
bonding states have mutually exclusive distance constraints that cannot be
satisfied simultaneously. In this event, the state is not allowed and must be
removed from the model. The layer simulation approach provides a reliable method
to eliminate states, since an estimate of the outer first passage time
$\tau_{(ij)}^+$ between states $i$ in layer $\ell$ and $j$ in layer $\ell + 1$
is computed before the entropy of state $j$. If there are no outer collisions in
all states $i$ that are connected to state $j$, and $\tau_{(ij)}^{+}$ is
infinite, state $j$ can be eliminated from the layer $\ell + 1$.

\begin{figure}[htbp]
    \label{sec:layerEfficiency}
    {\centering{
    \includegraphics[width=0.75\linewidth]{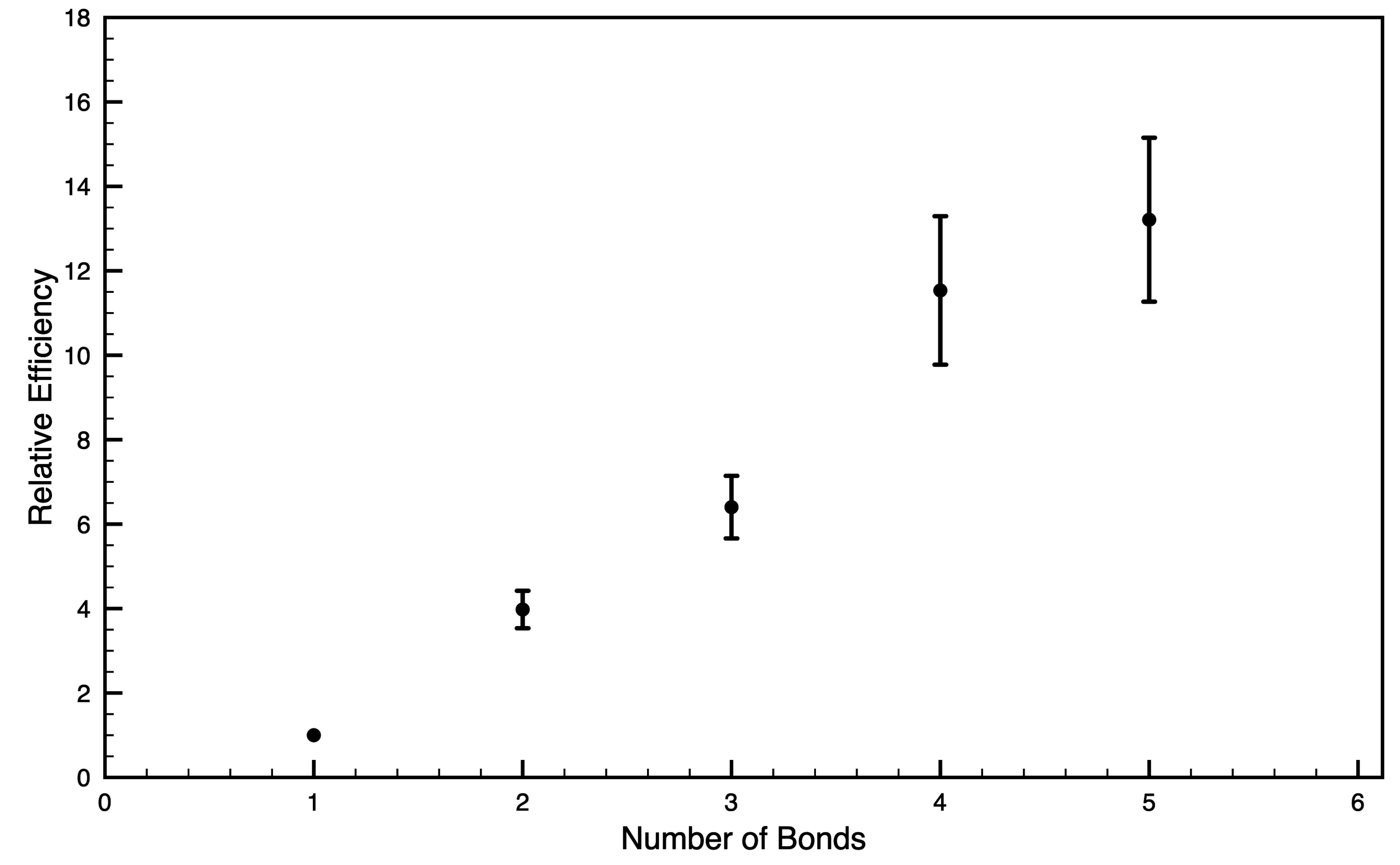}}}
    \caption{The ratio of the relative efficiency of the layers to the general
        procedure as a function of the number of bonds (or levels). In the
        general procedure, all states (rather than a pair) are accessible within
        a simulation. The relative efficiency is the ratio of the wall clock
        times needed to achieve convergence in each method. The data were
        generated by averaging the serial execution time to convergence of $100$
        instances of a $20$-bead model with $n_i = 400$ configurations sampled
    per state $i$ and a convergence level of $p = 0.25$.}
    \label{fig:efficiency}
\end{figure}

The decomposition of the calculation of the configurational entropy into a
number of independent calculations between adjacent layers reduces the
computational demands of the task relative to a procedure in which all nonlocal
bonds are active and $n_s = 2^{n_b}$ values of the entropy are evaluated
simultaneously. This gain in serial efficiency is due to the reduction in the
overall sampling time $\tau_s$ per sample required in the iterative procedure of
verifying the uniform convergence of the sampled states that scales as the
square of the number of bonds (see Sec.~(\ref{sec:layer})). In
Fig.~\ref{fig:efficiency}, the relative efficiency of the layer method is
demonstrated for a simple model with $20$ beads and a variable number of bonds
$n_b$. The two simulation approaches coincide for a model with a single bond
where $n_b = 1$ and $n_s = 2$, and it is evident that the relative efficiency of
the layer method increases roughly linearly with the number of bonds for a given
choice of sampled configurations per state.

\subsection{Biasing the entropy calculation: The staircase potential}
\label{sec:staircase}

\begin{figure}[htbp]
    \centering
    \includegraphics[width=0.75\linewidth]{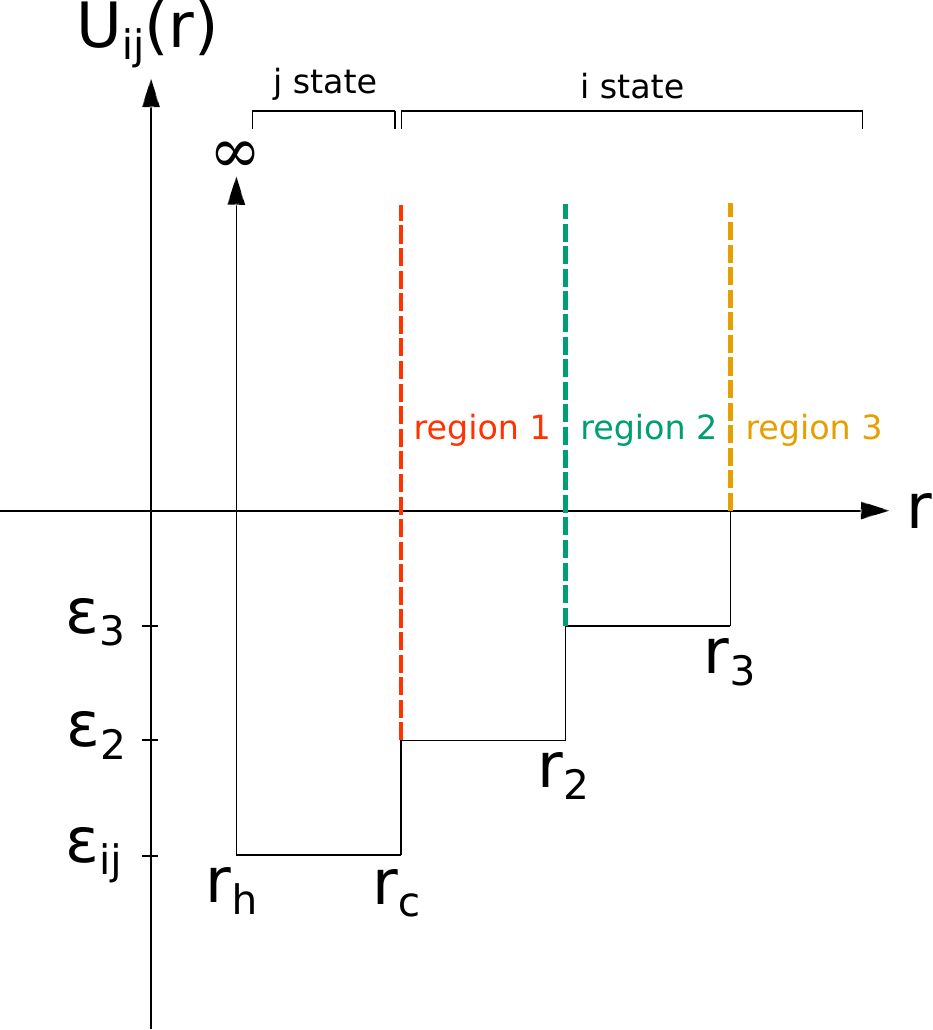}
    \caption{Example of a staircase with three steps. Note that $\epsilon_3 >
    \epsilon_2 > \epsilon_{ij}$ and $r_3 > r_2 > r_c$.}
    \label{fig:staircase}
\end{figure}

As the mean first passage time between two states increases, the length of the
trajectories $\tau_s$ required to sample independent configurations becomes
prohibitively large, rendering the direct calculation of the configurational
entropy difference between the states computationally inefficient. This
situation arises when the probability density for the bond distance
$\rho^{+}(r)$, which is the reaction coordinate for a change in state, is small
in the vicinity of the transition state at $r = r_c$. In the vicinity of $r_c$,
$C^{+}(r) \approx 0$ and the integrand in Eq.~\eqref{eq:mfpt+}, which is
proportional to $1/\rho^{+}(r)$, becomes large (see
Fig.~\ref{fig:staircaseFPT}). The probability density $\rho^{+}(r_c)$ at the
transition distance can be small either because i) in the unbonded state, the
range of the bond distance allowed is broad and the configurational volume of
the unbonded state is large or ii) existing nonlocal bonds in the initial state
introduce geometrical constraints in the chain that prevent the bonding distance
from being reached unless beads in the rest of the chain are placed optimally.
In both situations, the low probability of exploring the reaction coordinate
values in the region of the transition state results in inefficient sampling.
For systems with continuous potentials, sampling methods such as
metadynamics\cite{Laio:2002,Barducci:2008} or umbrella
sampling\cite{Marsili:2006} can be used to bias the stochastic sampling to visit
improbable regions of the reaction coordinate. However, event-driven dynamical
sampling is not amenable to the introduction of continuous force fields.

To encourage these bonds to form more readily while maintaining the
discontinuous nature of the model, we introduce modifications to the potential
described in Eq.~\eqref{eq:attractivestep} to bias the calculation of the
configurational entropy by reducing the required trajectory time $\tau_s$. This
bias is a computational device to calculate the biased entropy and mean first
passage times for the original model. Like an adaptive binning
strategy\cite{Bornn:2013}, we subdivide the outer region of the constant energy
potential to create a discontinuous potential that resembles a staircase, as in
Fig.~\ref{fig:staircase}.

To simulate dynamics in the staircase potential, we use a layer approach in
which each step of the staircase defines a new state that is implemented in a
separate layer. As a result of the restrictions to the bond distance, the mean
first passage time between the staircase regions is small, and only short
trajectories are required to evaluate the ratio of the relative number of
states.

When the initial state is divided into $\ell$ sub-states (such as those defined
by the $\ell = 3$ regions $1$, $2$ and $3$ in Fig.~\ref{fig:staircase}), the
configurational entropy for the $i$ to $j$ transition in the original model is
not the sum of the entropy differences between the regions due to the fact that
the total number of states in configuration $i$ is the sum of the number of {\it
states} in each of the regions. Rather, the entropy difference between states
$i$ and $j$ is
\begin{align}
    \label{eq:ientropy}
    e^{\Delta S_{ij}} &= \frac{n_i}{n_j} \\
    n_i &= \sum_{k=1}^{\ell} n_{ik}, \nonumber
\end{align}
where $n_i$ is the total volume of state $i$ and $n_{ik}$ is the volume of
region $k$. If each of the regions is treated as a separate layer,
$\tilde{S}_{k, k-1} = \ln \big( n_{ik}/n_{ik-1} \big)$ corresponds to the
entropy difference between the sub-states defined by adjacent regions $k$ and
$k-1$. If the volume of region $j$ is taken as $n_{i0}$, then we find
\begin{align}
    \label{eq:entropyStairs}
    \Delta S_{ij} = \ln \, \sum_{k=1}^{\ell} \exp \left\{\sum_{m=1}^{k}
    \tilde{S}_{m,m-1}\right\}.
\end{align}
Each value of $\tilde{S}_{m, m-1}$ is readily computed using the dynamical
sampling procedure in Sec.~(\ref{sec:layer}). Note that the estimate of the
entropy difference in Eq.~\eqref{eq:entropyStairs} has an approximate asymptotic
variance $4\ell / n_c$ when $\tilde{S}_{m, m-1} > 1$. Consequently, the number
of trajectories per iteration should be scaled appropriately for a given level
of precision.

The number of steps $\ell$ in the staircase potential and the location of each
of the steps $r_i$ can be estimated from the expected difference in entropy
$\Delta S_{ij}$ and the distribution of the relevant bond distance from the
simulation in the previous layer. If the drop in entropy in each of the regions
is constant, $\tilde{S}_{m,m-1} \approx \tilde{S}$, which means $\Delta S_{ij}
\approx \ell \tilde{S}$, and hence $\ell \approx \Delta S_{ij}/\tilde{S}$. The
location $r_i$ of staircase $i$ is determined from the cumulative bond distance
distribution $C^+(r)$ by the condition that $C^+(r_i) =
e^{-\tilde{S}}C^+(r_{i+1})$, where the outermost region satisfies $C^+(r_\ell) =
e^{-\tilde{S}}$.

We have found that for all models considered here, the largest change in entropy
$\Delta S_{ij} \approx 12$ so that a typical choice of $\tilde{S} = 4$ requires
the introduction of no more than three staircase regions. Larger choices of
$\tilde{S}$ result in less efficient sampling since the first passage time
between regions increases exponentially with $\tilde{S}$. For the special case
of $\ell = 3$, we have
\begin{align}
   \Delta S_{ij} = \tilde{S}_{1,j} &+ \tilde{S}_{2,1} + \tilde{S}_{3,2}
    \nonumber \\
    &+ \ln \left[1 + e^{-\tilde{S}_{3,2}} \left(1 +
    e^{-\tilde{S}_{2,1}}\right) \right].
\end{align}
For the crambin and frustrated models considered in Sec.~(\ref{sec:equildyn}),
the typical values of the step location were $r_2 \approx 1.8$ and $r_3 \approx
2.5$.

\begin{figure}[htbp]
    {\centering{
    \includegraphics[height=6cm]{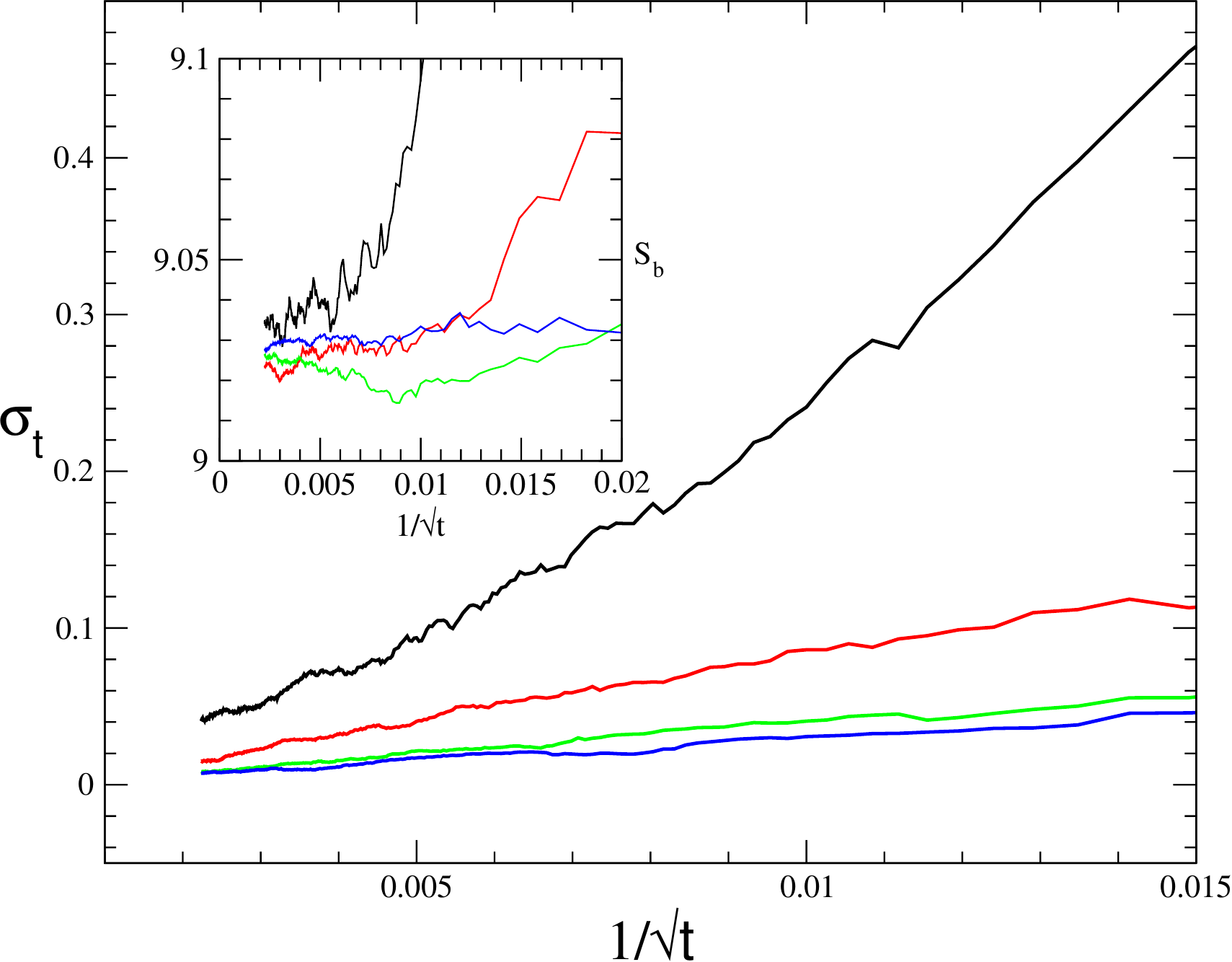}}}
    \caption{The standard deviation $\sigma_t$ of the drift term vs $1/\sqrt{t}$
        for a fixed scaled time step $t = n_s \tau$ averaged over $50$
        realizations of the adaptive procedure. The model is a $40$-bead system
        in which distant beads $10$ and $30$ form a nonlocal bond at $r_c =
        1.5$. The black line denotes a system with no staircase, and the red,
    green, and blue lines denote systems with one, two, and three additional
steps. The inset shows the value of $S_b(t)$ vs $1/\sqrt{t}$ for the
corresponding systems.}
    \label{fig:stairsVariance}
\end{figure}

The introduction of the staircase potential greatly reduces the workload of
computing the entropy difference between a pair of states that infrequently
interconvert, and it also improves the accuracy of the outer first passage
times. The staircase bias increases the rate of convergence of both the initial
Wang--Landau estimates of the entropy and the subsequent procedure for the
verification of convergence. To demonstrate this explicitly, we consider a
$40$-bead chain with a single nonlocal bond between bead pair $[10, 30]$. The
entropy difference between the non-bonded and bonded states is $S_0 - S_1 = 9.0
\pm 0.1$, and the mean first passage time for this model is approximately
$\tau^+ = 627 \pm 29$ due to the average separation between the bonding beads.
The convergence rate of the Wang--Landau procedure depends on the magnitude of
the standard deviation $\sigma_t$ of the drift term in the adaptive adjustments.
As is clear in Fig.~\ref{fig:stairsVariance}, the introduction of additional
staircase states into the system at a fixed computational cost reduces the
standard error of the Wang--Landau procedure. This error decreases as $t^{-1/2}$
with time step $t = n_s \tau$, even though the cost-per-iteration of the
algorithm increases linearly with the number of steps in the staircase. For this
model, the reduction saturates after the inclusion of two steps in the staircase
(green line in Fig.~\ref{fig:stairsVariance}). The ratio of the standard errors
$\sigma_i(t) \sim \sigma_i / \sqrt{t}$ for a simulation without a staircase to
one with $i$-steps is roughly $\sigma_0 / \sigma_3 \approx 23/3$, indicating
that the inclusion of the steps reduces the computational time needed at a given
level of statistical resolution by a factor of roughly $60$. At the same time,
the efficiency of the validation procedure to establish uniformity is also
improved, since the trajectory length required for each independent sample
decreases from $\tau_s \sim 8 \tau^+ / \pi \sim 6400$ trajectories of unit
length to less than $\tau_s \sim 50$.

The bias introduced by the stairs also improves the calculation of the outer
first passage time. The sampling of the reaction coordinate is enhanced in the
vicinity of the transition state at $r_c$, where the integrand of the
first passage time is the largest, as is apparent in Fig.~\ref{fig:staircaseFPT}
for the $40$-bead model. The density of the reaction coordinate $\rho^+(r)$ is
constructed by stitching together continuous fits of the densities in each of
the staircase regions. To improve the quality of the fit of the integrand in the
region near $r_c$, a larger number of sampling points in the staircase region
containing $r_c$ should be used. Without enhancing the sampling, the standard
error of the estimated outer first passage time is large when transitions are
rare.

\section{Folding dynamics, pathways, and evolution} \label{sec:equildyn}

The simplicity of the discontinuous potential model allows both the free energy
and the transition rate matrix $\bm{K}$ in a Markovian description of the
dynamics to be determined analytically for any choice of state energies at any
temperature. These features enable the study of how folding pathways from a
non-bonded initial state to the fully bonded ``folded" state change with these
parameters.

The utility of Markov chains in describing the dynamics of chemical and
biophysical systems has long been recognized, and vast literature exists on the
subject (for example, see Refs.~\onlinecite{vanKampen:2007,Allen:2003}). A
number of properties are of interest in a Markov state model of protein
dynamics. Since $\bm{K}$ is a regular, square matrix satisfying detailed
balance, it has a unique zero eigenvector that corresponds to the equilibrium
populations. The transition rate matrix can be written in terms of a diagonal
matrix as $\vect{K} = \vect{U} \lambda \vect{U}^{-1}$, where $\vect{U}$ is a
matrix with eigenvectors of $\vect{K}$ as the columns, and $\lambda$ is a
diagonal matrix with eigenvalues $\lambda_i \leq 0$ on the diagonal. The
spectrum of eigenvalues $\{\lambda_i\}$ can be useful to determine if a small
number of states dominate the long-time dynamics of the system. When this is the
case, reduction techniques such as stochastic complementation may be profitably
applied to reduce the dimensionality of the Markov model\cite{Meyer:1989}. In
addition, the probability of particular paths starting from an initial
distribution of states to the folded state can be analyzed to find dominant
folding pathways and potential bottlenecks in the non-equilibrium first passage
path ensemble\cite{vonKleist:2018,Sharpe:2021}.

Functional proteins have evolved to carry out specific tasks under stressful
environmental conditions. Since their function is intimately linked to their
three-dimensional structure, their structure must be resilient to thermal
stress. This suggests that a fast-folding, single-domain protein should not only
exhibit a strong preference for its active structure over a range of
temperatures, but it should also rapidly equilibrate or refold to this ``native"
structure if perturbed. Naturally evolved proteins of this type have optimized
sequences and energies of configurations that result in such characteristics.

The evolution of sequences selected to optimize thermal stability can be
examined in the Markov state model by considering the variation of the folding
time with respect to the set of interactions $\{\beta^* \vect{E}\} = \{
\vect{E}^*\}$ in the model. The folding time can be analyzed by considering the
probability density $S(t)$ of the system in a non-native configuration at time
$t$ in the presence of an absorbing state $f$, defined as
\begin{equation}
    S(t) = \sum_{i \neq f} P_i(t) \nonumber.
\end{equation}
Here $f$ is taken to be the index of the folded (native) state and $P_i(t)$ is
the population of state $i$ at time $t$. Assuming the initial state of the
system is the fully unfolded state of index $u$, $S(t)$ can be written for the
Markov state model as
\begin{equation}
    S(t) = \sum_{i=1}^{n_s-1} \left(e^{\vect{\tilde{K}} t}\right)_{iu},
    \nonumber
\end{equation}
where $\vect{\tilde{K}}$ is the square matrix of rank $n_s - 1$ obtained by
removing the row and column from the transition matrix $\vect{K}$ corresponding
to the native state $f$. This matrix is invertible and has negative real
eigenvalues.

The first passage time density $f(t)$ to the folded state is
\begin{equation}
    f(t) = -\frac{dS(t)}{dt} = -\sum_{i \neq f} \frac{dP_i(t)}{dt}, \nonumber
\end{equation}
and hence the mean and variance of the folding time are given by
\begin{align}
    \mu_t (\{\vect{E}^*\}) &= \int_0^{\infty} t f(t) \, dt = -\sum_{i=1}^{n_s-1}
    \vect{\tilde{K}}^{-1}_{iu} \label{eq:mean_fpt} \\
    \sigma_t^2 (\{\vect{E}^*\}) &= \sum_{i, j=1}^{n_s -1}
    \left(2\,\vect{\tilde{K}}^{-1}_{ij} \vect{\tilde{K}}^{-1}_{ju} -
    \vect{\tilde{K}}^{-1}_{iu} \vect{\tilde{K}}^{-1}_{ju}\right),
    \label{eq:variance_fpt}
\end{align}
which depend on the choice of the set $\{\vect{E}^*\}$ of dimensionless
interaction energies $E_i^* = \beta^* E_i$.

There are a number of dynamical measures that are helpful to understand the
characteristic behavior of a Markov state model. We consider an ensemble of
``reactive" trajectories defined as the set of trajectories initiated from the
unfolded state $u$ that reach the folded state $f$ without revisiting the
initial state\cite{vonKleist:2018}. The definition of the ensemble makes use of
the committor probability $q_i^+$ that a trajectory from a given state $i$
reaches the folded target state $f$ before reaching the unfolded state $u$,
\begin{equation}
    q_{i}^+ = -\sum_{j \neq (u, f)} \vect{K}_{fj} \,
    \dbtilde{\vect{K}}^{-1}_{ji}, \label{eq:committor}
\end{equation}
where the matrix $\dbtilde{\vect{K}}$ is obtained from $\vect{K}$ by removing
the rows and columns of the $u$ and $f$ states. We assume that set of
macrostates defining the originating set in the reactive ensemble consists only
of the unfolded state $u$ and that the committors $q^{+}_{i}$ are non-zero for
$i \neq u$. One defines the transition probability matrix $\vect{T}$ of passing
from state $i$ to state $j$ from the transition rate matrix $\vect{K}$
as\cite{vonKleist:2018,Sharpe:2021}
\begin{equation}
    T_{ji} = \frac{K_{ji}}{\sum_{k \neq i} K_{ki}},
\end{equation}
and the reactive transition matrix $\tilde{\vect{T}}$ with elements
$\tilde{T}_{jf} = \delta_{j,f}$, $\tilde{T}_{u,i} = 0$ as
\begin{align}
    \tilde{T}_{ji} =
    \begin{dcases}
        \frac{q_j^+ T_{ji}}{q_i^+} & \text{if $i\neq u$, $i \neq f$} \\
        \frac{q_j^+ T_{ju}}{\sum_{j \neq u} q_j^+ T_{ju}} & \text{if $i=u$}.
    \end{dcases}
\end{align}
The elements $\tilde{N}_{ji}$ of the fundamental matrix $\tilde{\vect{N}} =
(\vect{I} - \tilde{\vect{T}})^{-1}$ are the expected number of visits to state
$j$ from state $i$ in the reactive ensemble. The expected number of visits
$\tilde{\theta}_j$ for any state $j$ from an ensemble of reactive trajectories
initiated from the unfolded state $u$ is given by $\tilde{\theta}_j =
\tilde{N}_{ju}$. Similarly, the visitation probability matrix is denoted as
$\tilde{\vect{H}}$, and it satisfies $\tilde{\vect{N}} = \vect{I} +
\tilde{\vect{H}} \cdot \tilde{\vect{N}}$. Its elements $\tilde{H}_{ji}$
correspond to the probability that a reactive trajectory initiated at state $i$
will reach state $j$, where the reactive probability
\begin{equation}
    r^+_j = \tilde{H}_{ju} \label{eq:reactiveProbability}
\end{equation}
is the probability that state $j$ will be visited along the reactive path,
starting from the unfolded state\cite{Sharpe:2021}.

The reactive flux, $\tilde{J}_{ji}$, measuring the reactive rate from state $i$
to state $j$, is defined as\cite{vonKleist:2018}
\begin{equation}
    \tilde{J}_{ji} = \left(\frac{q_j^+ T_{ji}}{\sum_k q_k^+ T_{ki}}\right) \,
    \tilde{\theta}_i. \label{eq:reactiveFlux}
\end{equation}
The reactive probability $r_j^+$ and the reactive fluxes $\tilde{\vect{J}}$
provide useful measures of the probability of different pathways, and the
importance of a particular state in the folding process. We make use of these
quantities in Sec.~\ref{sec:crambin} and Sec.~\ref{sec:frustratedModel}.

\begin{figure*}[htbp]
    \hfill
    \subfloat[
        \label{fig:crambinPDB}
        Crambin crystal structure from the Protein Data Bank
    ]{%
        \includegraphics[height=3.5cm]{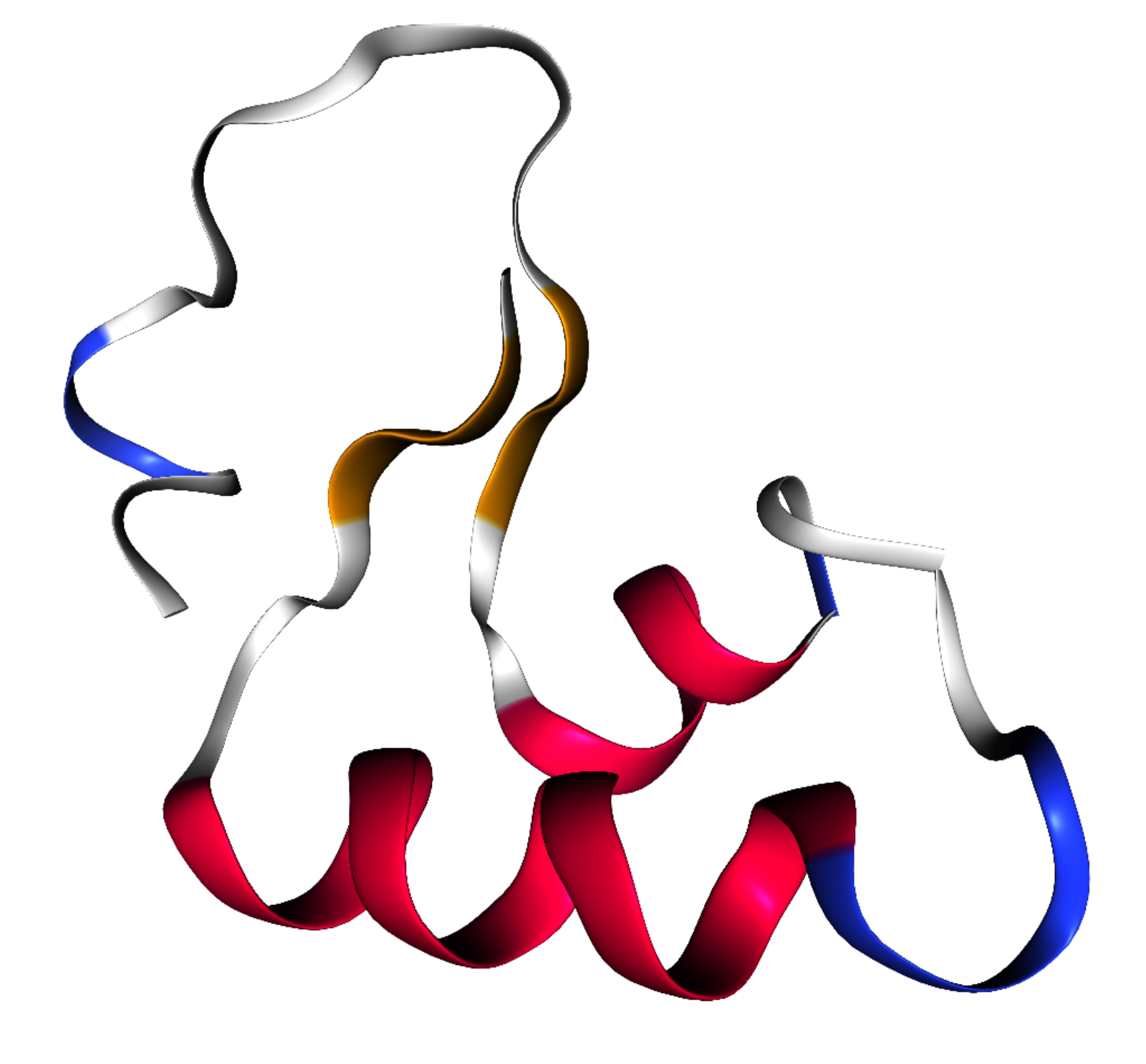}%
    }\hspace{25ex}
    \subfloat[
        \label{fig:crambinNative}
        Native state of the crambin model: state $1024$
    ]{%
        \includegraphics[height=3.5cm]{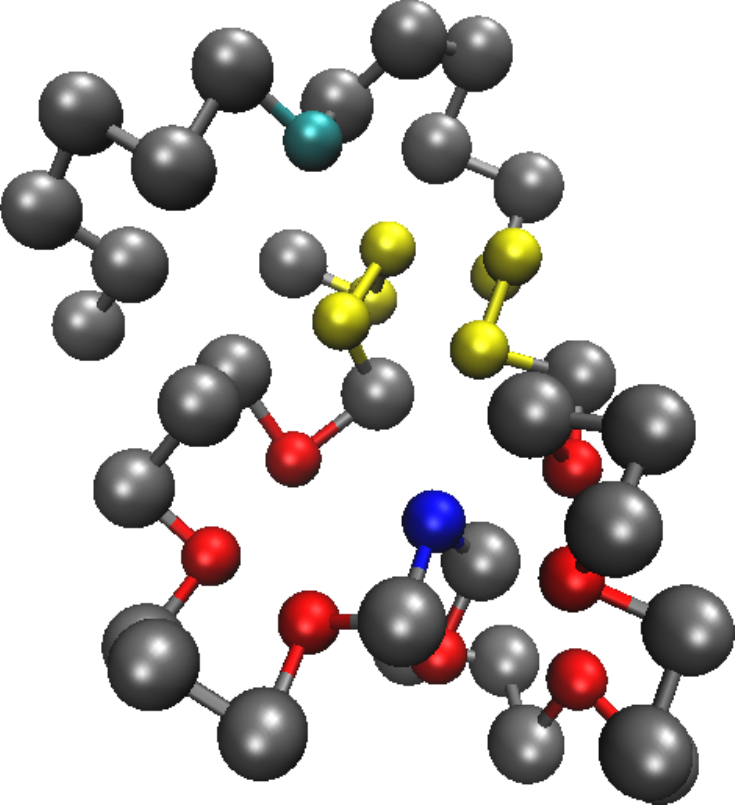}%
    }\hspace{25ex}
    \subfloat[
        \label{fig:crambinTransition}
        Transition state $617$
    ]{%
        \includegraphics[height=3.5cm]{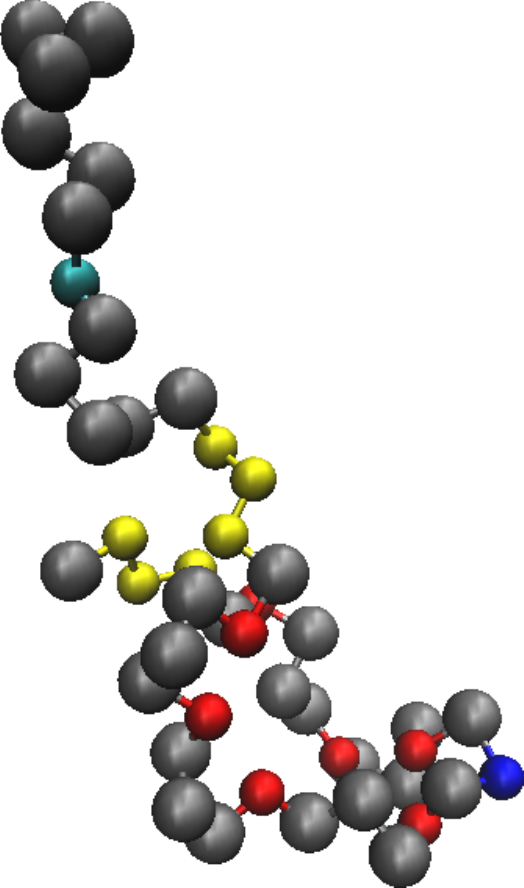}%
    }
    \hfill
    \label{fig:crambinModel}
    \caption{The model crambin system. Fig.~\ref{fig:crambinPDB} is a cartoon
        representation of the structure of the crystallized protein.
        Fig.~\ref{fig:crambinNative} is the fully folded minimum entropy state
        of the $46$-bead, $10$-bond model. The structure in
        Fig.~\ref{fig:crambinTransition} contains five bonds leading to a
        helical structure but none of the four nonlocal bonds between distant
        monomers that culminate the folding process. In all three figures, the
        beads participating in nonlocal bonds in the $\alpha$-helices are in
    red, the $\beta$-sheets are in yellow, and the disulfide bridges are in blue
and cyan for beads $16$ and $40$, respectively.}
\end{figure*}

The evolutionary process for the model system can be simulated by defining a set
of beneficial physical characteristics that the system should have. For real
biological systems, the selection pressures vary according to their environment
and the required physical function of the biomolecule. The relevant physical
characteristics, which depend on the set of state energies $\{E_i^* | \, i=1,
\dots, n_s\}$, could include the requirement that the native state is the most
probable state of the system over a large range of temperatures. Additionally,
interactions can be selected that make both the mean folding time $\mu_t$, given
in Eq.~\eqref{eq:mean_fpt}, and the variance $\sigma_t^2$ in
Eq.~\eqref{eq:variance_fpt}, as small as possible assuming a fixed ratio of the
probabilities of the unfolded state to folded state, $P_u / P_f$. These
constraints on the selection of energies ensure that the protein not only folds
and refolds quickly, but is also unlikely to have folding pathways that trap
intermediate structures for extended periods of time.

Here, we consider the simple loss function for the constrained variational
optimization of the set of energies $\{\vect{E}^*\}$,
\begin{equation}
    {\cal L} = \mu_t(\{\vect{E}^*\}), \label{eq:lossFunction}
\end{equation}
to be the average folding time given in Eq.~\eqref{eq:mean_fpt} from the
non-bonded state $u$ to the folded state $f$. The most favorable choice of
interaction energies for a fixed ratio of $P_u / P_f$ is determined by
minimizing the loss function ${\cal L}$ with respect to the $n_s - 2$ adjustable
interaction energies. Additionally, the optimization is constrained such that no
intermediate state $i$ is substantially populated by including the inequality
condition $P_i / P_f \leq 0.005$ to maintain the dominance of the native
population. Many choices of selective pressure, defined by the loss function and
constraints, are possible and relevant for other types of proteins with
different functionality. Note that the gradients of the loss function can also
be computed analytically from $\vect{\tilde{K}}$, and the probabilities of the
configurations can be used to accelerate the minimization of the loss function.
For the simple loss function in Eq.~\eqref{eq:lossFunction}, the gradients are
given by
\begin{equation}
    \frac{\partial \mu_t}{\partial E_k^*} = \sum_{\ell, m, n=1}^{n_{s}-1}
    \tilde{K}^{-1}_{\ell m} \frac{\partial \tilde{K}_{mn}}{\partial E^*_k}
    \tilde{K}^{-1}_{nu},
\end{equation}
and the derivatives of the $\vect{K}$ matrix are
\begin{align}
    \frac{\partial K_{ji}}{\partial E_k^*} &=
    \begin{dcases}
        \displaystyle{K_{ji}^2 \, \tau^{-}_{(ij)}
        \frac{P_i}{P_j}\left(\delta_{i,k} - \delta_{j,k}\right)} & j > i \\
        \displaystyle{K_{ji}^2 \, \tau^{+}_{(ij)}
        \frac{P_i}{P_j}\left(\delta_{i,k} - \delta_{j,k}\right)} & i > j \\
        -\displaystyle{\sum_{l \neq i} \frac{\partial K_{li}}{\partial E_k^*}} &
        i = j,
    \end{dcases}
\end{align}
where states are ordered by their number of bonds from fewest to most. The
derivatives of the matrix $\tilde{\vect{K}}$ or $\dbtilde{\vect{K}}$ can be
obtained from the derivatives of $\vect{K}$ by the appropriate removal of the
rows and columns at the index of the absorbing and source states. The numerical
minimization of Eq.~\eqref{eq:lossFunction} is complicated by the high-dimension
of the parameter search. Standard minimization algorithms, such as the
Nelder--Mead or Broyden--Fletcher--Goldfarb--Shanno (BFGS) methods, make use of
gradients, but this can lead to difficulties in locating the global minimum. For
such situations, methods that combine local gradient search algorithms with
multiple trajectory sampling are suitable\cite{Tseng:2008,Liao:2011,Kumar:2015}.

\subsection{The structure and dynamics of crambin} \label{sec:crambin}

\begin{figure*}[htbp]
    \subfloat[
        \label{fig:crambinTree}
        Tree diagram: crambin model
    ]{%
        \includegraphics[width=0.45\textwidth]{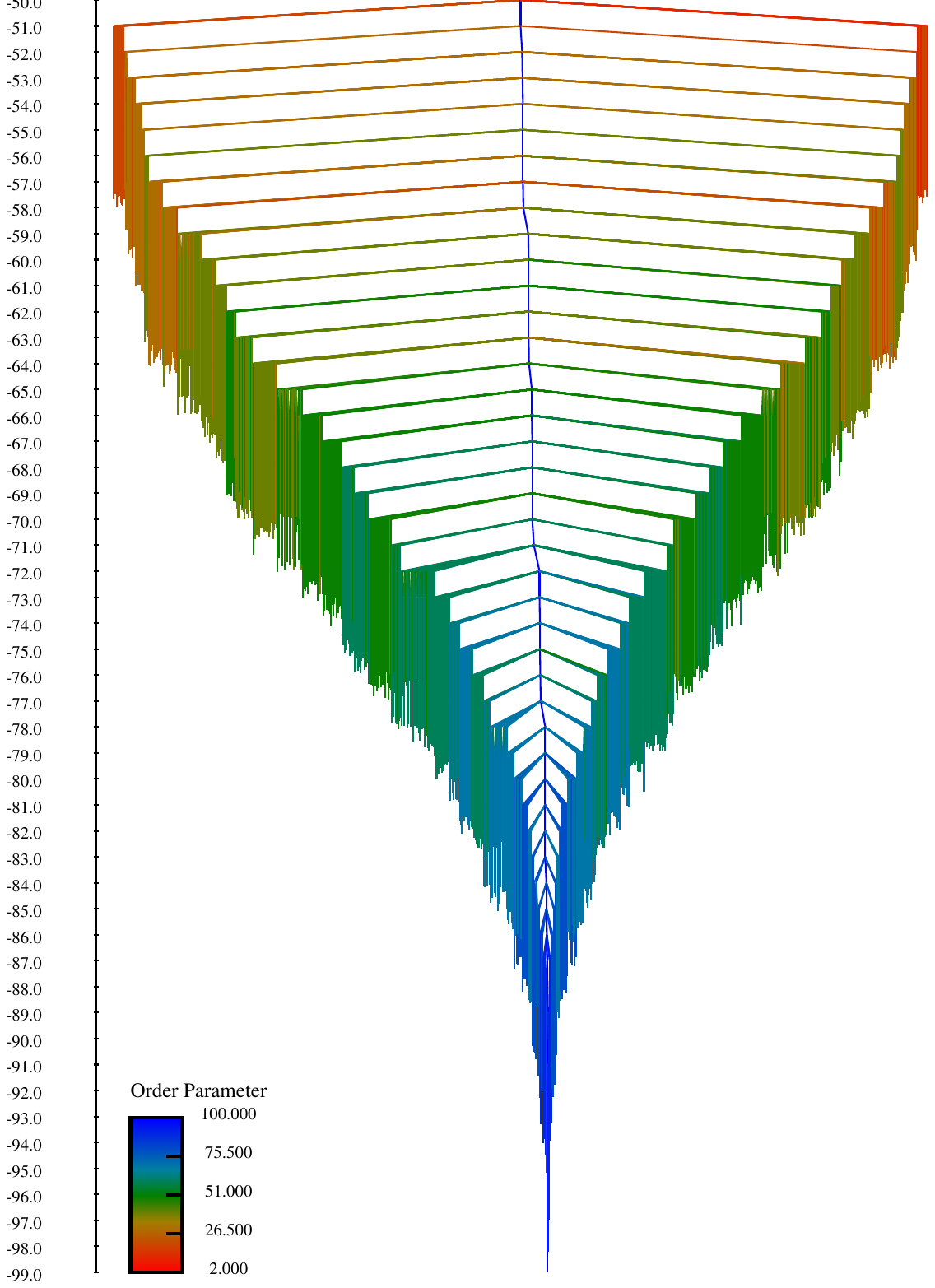}%
    }\hfill
    \subfloat[
        \label{fig:optimizedcrambinTree}
        Tree diagram: optimized crambin model
    ]{%
        \includegraphics[width=0.45\textwidth]{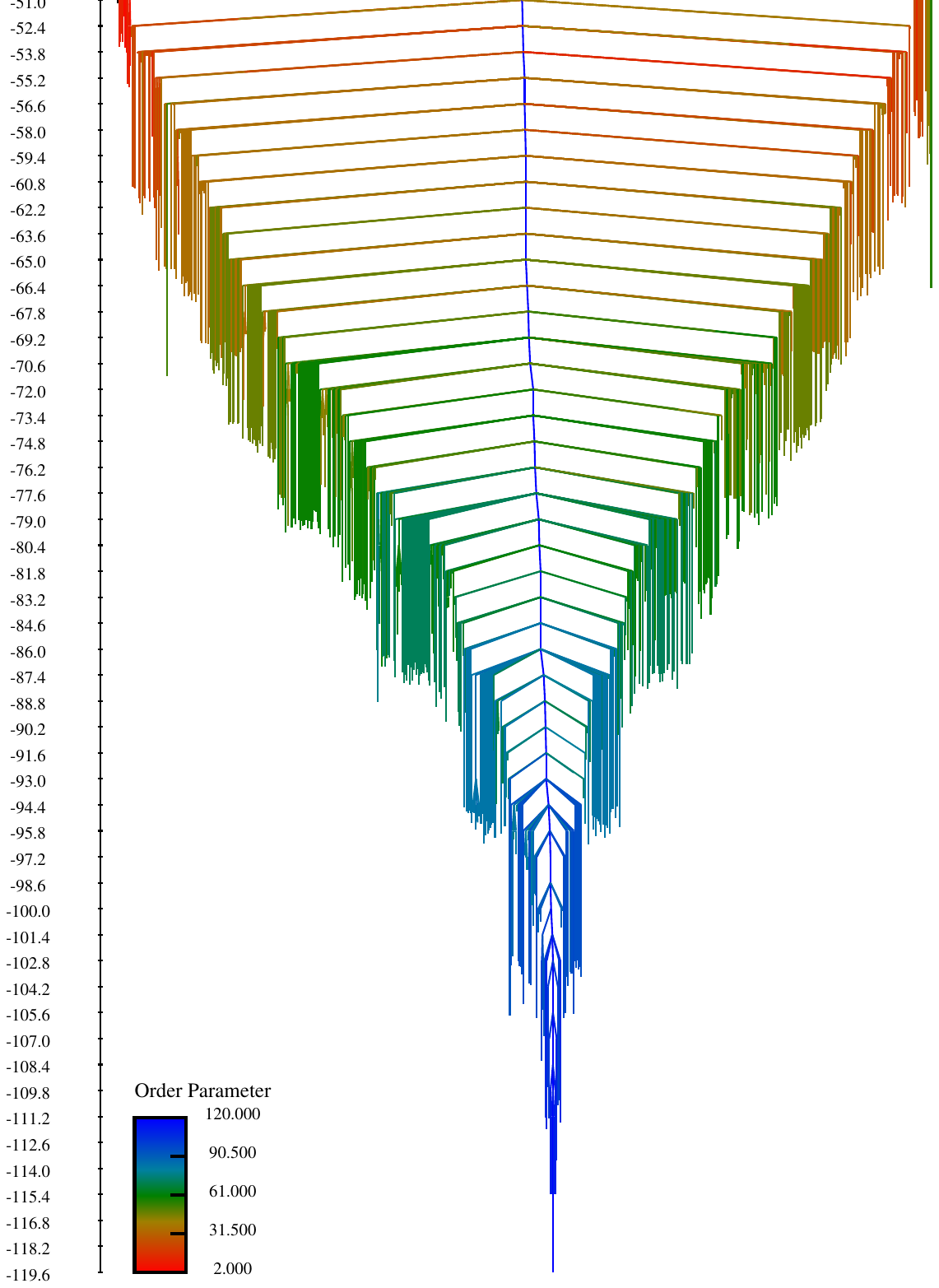}%
    }
    \caption{The disconnectivity graphs for the model crambin system in the low
    temperature regime, with $\beta = 12$.}
\end{figure*}

\begin{figure}[htbp]
    {\centering\includegraphics[height=5.5cm]{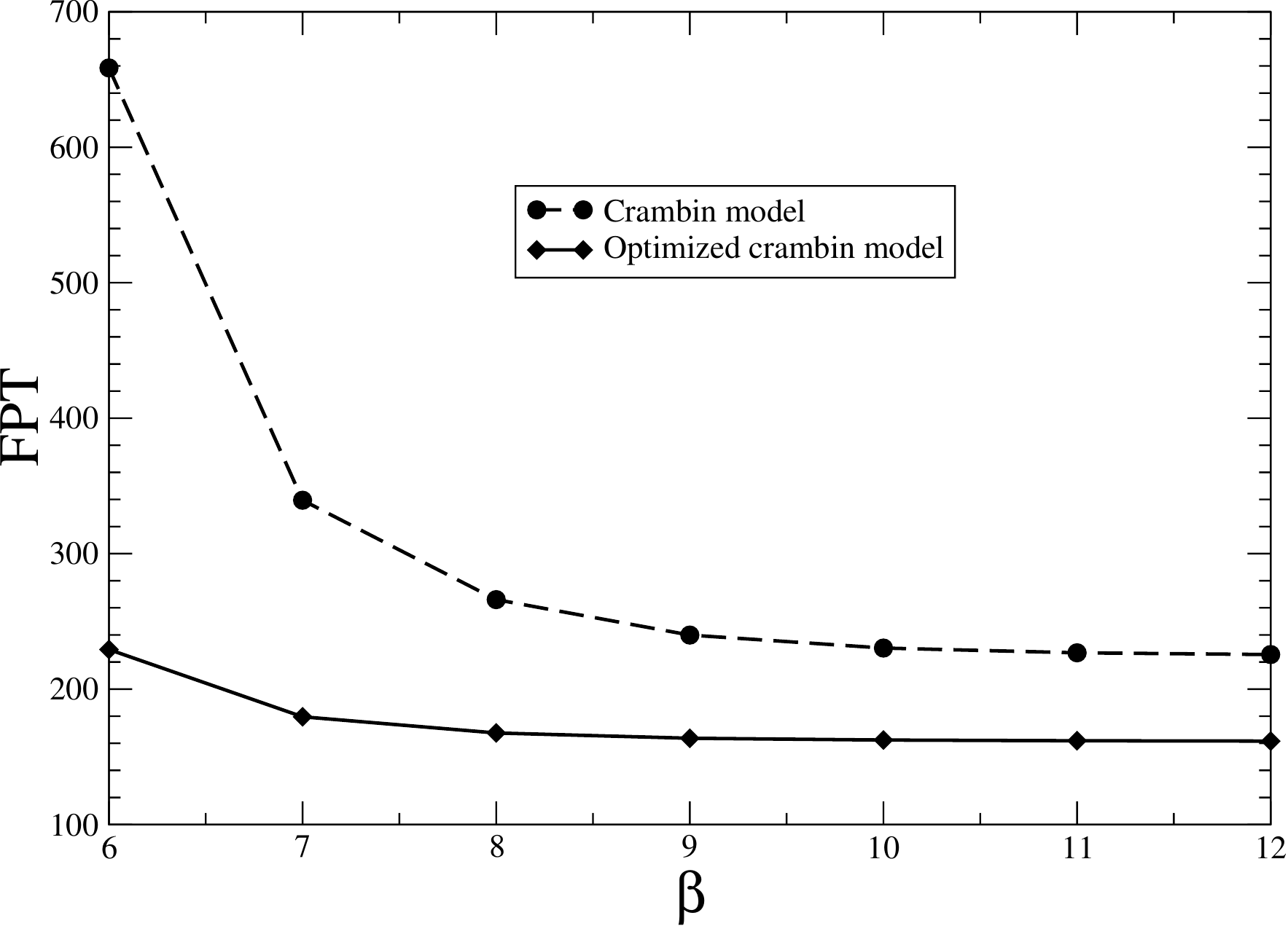}}
    \caption{The average folding times in units of $1 / D$ as a function of
    inverse temperature $\beta$ for the crambin model.}
    \label{fig:crambinFoldingTimes}
\end{figure}

\begin{figure}[htbp]
    \centering
    \subfloat[
        \label{fig:frustratedNetwork}
        Network diagram: equal bond energy crambin model
    ]{%
        \includegraphics[height=5.5cm, width=0.48\textwidth]{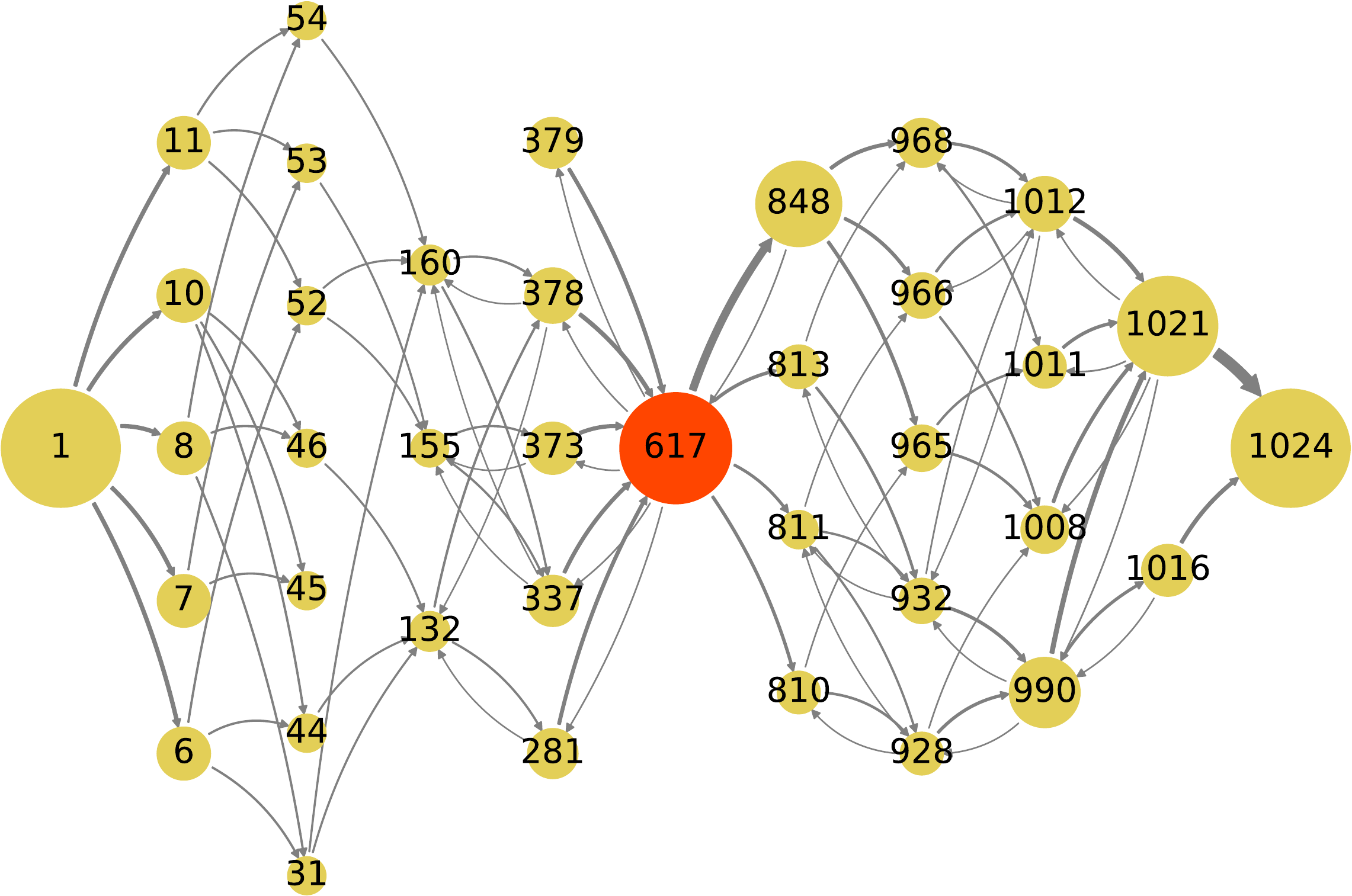}%
    }\hfill
    \newline
    \subfloat[
        \label{fig:optimizedNetwork}
        Network diagram: optimized energy crambin model
    ]{%
        \includegraphics[height=5.5cm, width=0.48\textwidth]{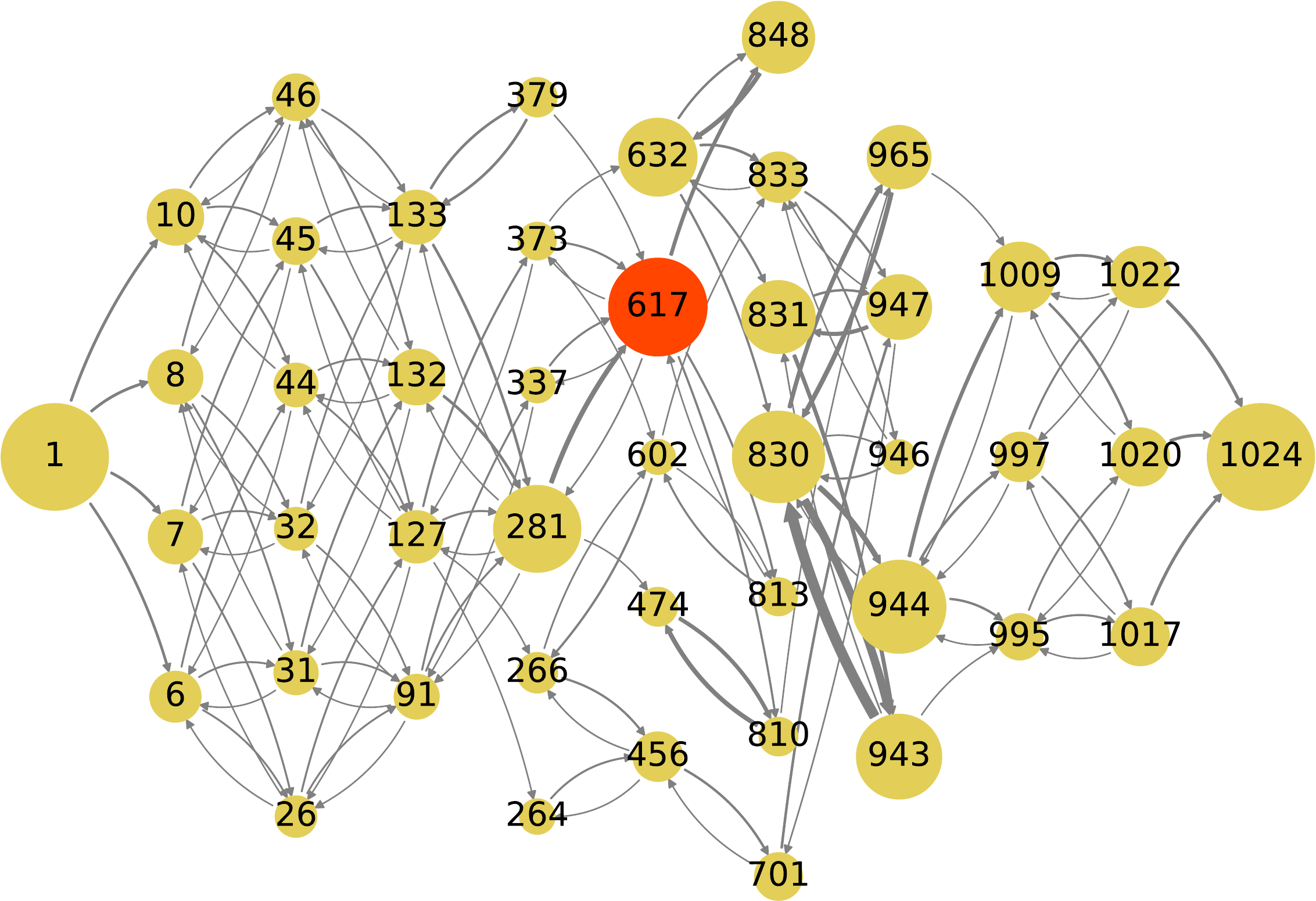}%
    }
    \caption{Simplified network diagrams of the most probable network of folding
        pathways for the model crambin system at low temperatures, where
        $P_u/P_f = 10^{-52}$, and in which only states with reactive probability
        $r^+ > 0.1$ are shown (see Eq.~\eqref{eq:reactiveProbability}). The size
        of each node is representative of the probability of visiting the state
        in the reactive ensemble, and the size of the arrows between nodes
        represents the reactive flux between them (see
        Eq.~\eqref{eq:reactiveFlux}). On the left is the network for a system
        with bonds of equal energy with bottleneck state $617$ colored red, and
        on the right is the network following the optimization of the folding
        time. Note that the two models have very different networks and folding
    pathways.}
\end{figure}

We now examine a coarse-grained model of crambin (PDB ID:
1EJG)\cite{Jelsch:2000,Jelsch:2000/3171}, a $46$-residue protein of unknown
function found naturally in cabbage. The three-dimensional crystal structure of
the protein, shown in Fig.~\ref{fig:crambinPDB}, has been measured with $0.48$
\angstrom resolution with x-ray crystallography\cite{Schmidt:2011}. The
structure of crambin is interesting, since it contains three important
structural motifs: $\alpha$-helices (in red), anti-parallel $\beta$-sheets (in
yellow), and disulfide bridges (in blue). The discontinuous model of crambin was
constructed from the crystal structure in the Protein Data Bank. To model
$\alpha$-helices, attractive interactions representing hydrogen bonds occur
only between nonlocal beads whose indices are $a = 2 + 4k$ and $b = a + 4l$,
where $l = 1, 4, 5 \dots$ and $k$ is any positive
integer\cite{Bayat:2012/245103}. The omission of bonds between monomers
separated by eight or twelve beads is done to discourage the formation of turns
and introduce rigidity along the protein's primary structure to prevent it from
collapsing in on itself over short distances. For other nonlocal interactions,
bonds were identified based on distances in the crystal structure, an idea used
in the construction of elastic network models\cite{Tirion:1996}. In particular,
crambin was assumed to have bonds formed at a distance $r_c = 1.5$ between a set
of beads separated by four residues, viz., $\{[6, 10], [10, 14], [14, 18]\}$ and
$\{[22, 26], [26, 30]\}$, that form two short $\alpha$-helices. The relative
orientation of the helices is restricted by an anti-parallel $\beta$-sheet
formed by bonds $\{[2, 34], [3, 33]\}$. The $\beta$-sheets are linked to the
terminal section of the protein, which has a random coil nature, by disulfide
bridges $\{[3, 40], [4, 32]\}$, and an additional bridge bond $[16, 26]$ links
the helices. The resulting ``native" structure when all bonds are formed is
shown in Fig.~\ref{fig:crambinNative}.

The evaluation of the drop in entropy and the outer first passage times for the
formation of the long-range disulfide bonds benefits from the use of the
staircase bias, given that the decrease in entropy for such bonds is roughly
$10$ and $\tau^+ \sim 10^3$. The input parameters, mean first passage times, and
biased entropies for the simulation of crambin are available on GitHub (see
Acknowledgements) in the hybridmc/examples folder. To visualize the free energy
landscape and the kinetics for a pairwise-additive model in which each bond
formed lowers the potential energy by a fixed amount $\epsilon_b$, we use
disconnectivity graphs\cite{Becker:1997,Wales:2004}. The node levels in the
graphs are determined by the dimensionless free energies, and the transition
state free energies are set by adding the negative logarithm of the rate to the
state's free energy. Changes in the morphology of the free energy landscape as
the temperature is modified can be tracked by the structure of the
disconnectivity graphs. Unsurprisingly, the disconnectivity graph for the model
crambin system shown in Fig.~\ref{fig:crambinTree} exhibits a ``funnel-shape" at
low temperatures ($\beta^*= 12$), in which the fully bonded structure
corresponds to a deep-lying node in the graph, centrally flanked by local minima
over a wide range of free energy values\cite{Becker:1997}. The folding dynamics
for the model, which exhibits no kinetic traps, is particularly simple. The
average folding time needed to pass from an initial state with no bonds to the
target native state, shown in Fig.~\ref{fig:crambinFoldingTimes} in
dimensionless units inversely proportional to the self-diffusion coefficient
$D$, decreases monotonically as $\beta$ increases and approaches a constant
value. Note that this does not imply that the folding rate is fastest at low
temperatures, since from kinetic theory\cite{Chapman:1990}, the diffusion
coefficient is expected to scale as $D \sim \beta^{-1/2}$ so that the physical
folding time increases at low temperatures.

The most probable pathways of transitioning from the non-bonded state to the
folded target state can be visualized using network diagrams, in which each
state appears as a node whose size is represented by the probability $r^+$ (see
Eq.~\eqref{eq:reactiveProbability}) that the state is visited in the reactive
ensemble. The connecting arrows represent the reactive flux (see
Eq.~\eqref{eq:reactiveFlux}). For the equal bond energy model at low
temperatures where the rate of breaking a bond is small and the committor
probability $q^+_i \sim 1$ for all bonded states, the folding pathways primarily
consist of two distinct parts: Five local bonds are formed first, leading to a
helical intermediate state $617$ (colored red in
Fig.~\ref{fig:crambinTransition}, with $\beta$-sheet $[2, 34]$, disulfide bridge
$[4, 32]$, and $\alpha$-helix bonds $[6, 10], [14, 18], [26, 30]$ turned on),
followed by the formation of the disulfide bridge $[16, 26]$, the $\beta$-sheet
$[2, 34], [3, 33]$, and the $[4, 32]$ bond. In $70\%$ of the folding pathways,
the most probable final transition to the folded structure involves the
formation of the disulfide bridge $[3, 40]$, linking the $\beta$-sheet to the
random coil end of the chain, denoted as state $1021$ to state $1024$.

Heretofore, we have assumed that the formation of a bond changes the energy of a
configuration by an amount $\epsilon_b$. Suppose we are interested in
determining the optimal set of interactions that lead to a given structure,
while maintaining a set of physical requirements. Namely, the fully bonded
structure has a free energy that is well-separated from other structures so that
it is thermodynamically preferred over a range of temperatures, yet is reached
quickly from a fully unbonded configuration. In principle, since the
coarse-grained models are allowed a nonadditive (i.e., not pairwise)
decomposition of the potential energy to permit hidden effects such as
hydrophobicity not directly incorporated into the model (see
Eq.~\ref{eq:attractivestep}), arbitrary choices of the energies of states are
possible provided they are physical. To mimic evolutionary behavior, we minimize
the mean folding time in Eq.~\eqref{eq:lossFunction} with respect to the set of
state energies $\{\vect{E}^*\}$, subject to the constraints that 1) the ratio of
the probability of the unfolded state to the folded state is fixed (i.e., the
state energies of the folded and unfolded states are constant), 2) the
probability that each partially folded state cannot be too large, enforced by a
constraint $0.005 < P_i / P_f$, and 3) the maximum energy of a given state is
restricted to a finite value (taken here to be less than $10$, well above the
zero energy of the unbonded state). For the crambin model with ten bonds, there
are $1022$ intermediate states whose energies are varied to minimize the mean
folding time. To carry out the minimization procedure of a loss function with
many possible local minima, we use methods that combine the BFGS search
algorithms with multiple trajectory
sampling\cite{Tseng:2008,Liao:2011,Kumar:2015}.

The result of the minimization procedure with $P_u/P_f = 10^{-52}$, a value of
the relative probability corresponding to $\beta^* = 12$ when the bond energy is
fixed at $\epsilon_b = 1$, lowers the mean folding time by a factor of roughly
$2$ over a range of temperature values, as shown in
Fig.~\ref{fig:crambinFoldingTimes}. Nonetheless, the smooth funnel morphology of
the disconnectivity graph is maintained (see
Fig.~\ref{fig:optimizedcrambinTree}). The disconnectivity graph of the optimized
model is more segmented, particularly in the last level of states, with most
states at a given level having similar probability and hence roughly the same
free energy. From the network diagram of the optimized model shown in
Fig.~\ref{fig:optimizedNetwork}, it is apparent that the folding mechanism is
significantly altered. The optimization yields energies of states that make the
pathways leading to the helical transition state equally likely (similar values
of $r^{+}$), and the effect of bottleneck state $617$ is mitigated by
substantially facilitating the $\beta$-sheet formation by decreasing the
energies of states with long-range bonds (such as the $\beta$-sheet $[2, 34]$
and $[3, 33]$ bonds) to allow additional connecting pathways at level $5$.
These findings are consistent with the view that structure grows locally and
models with local stabilizing interactions that compensate the conformational
entropy loss as local structure forms result in faster folding\cite{Munoz:1999}.
The change in folding mechanism and the increase in the folding rate correlate
with the increase in the ``contact order" in which the mean separation in
sequence between bonding beads\cite{Plaxco:1998,Dinner:2001} is weighted by the
reactive probability $r^+$ for that bond. There are also rapid transitions
between the state $830$ and $943$ that both have the $[3, 33]$ bond and then
form or break the adjacent $\beta$-sheet $[2, 34]$ bond. The energies of states
in the penultimate level are similarly adjusted to create three equally-likely
pathways to the final state. These results imply that even a system with a
smooth funnel will fold more quickly when the state energies allow for a
multitude of pathways rather than passing through a fixed sequence of states, in
agreement with studies of fast-folding proteins\cite{Wolynes:1995/1619}.

\subsection{Eliminating frustration and misfolding} \label{sec:frustratedModel}

\begin{figure*}[htbp]
    \subfloat[
        \label{fig:frustratedNative}
        Fully-bonded state: state $37$
    ]{%
        \centering\includegraphics[width=0.3\textwidth]{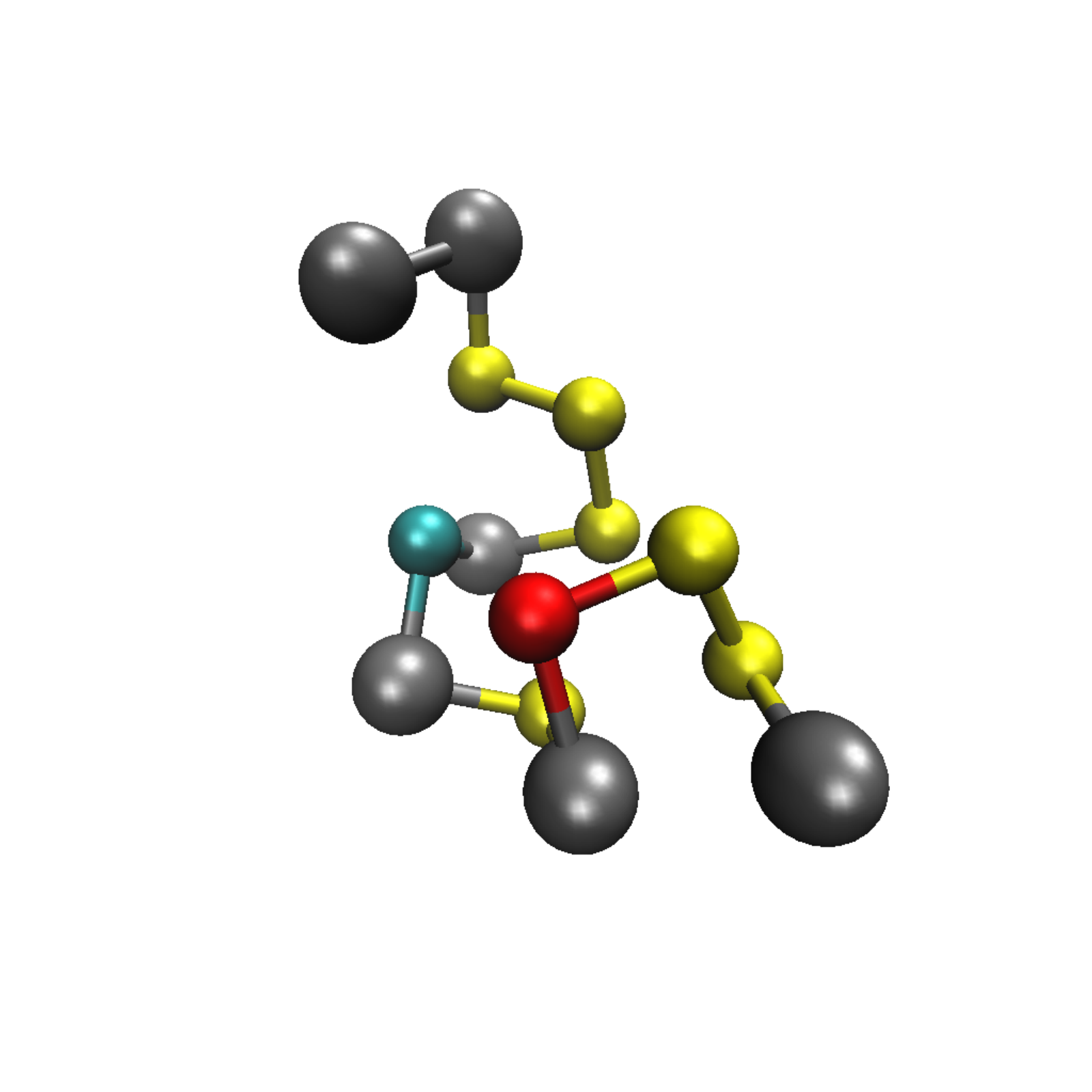}%
    }\hfill
    \subfloat[
        \label{fig:frustratedTransition}
        Maximum flux state: state $36$
    ]{%
        \includegraphics[width=0.3\textwidth]{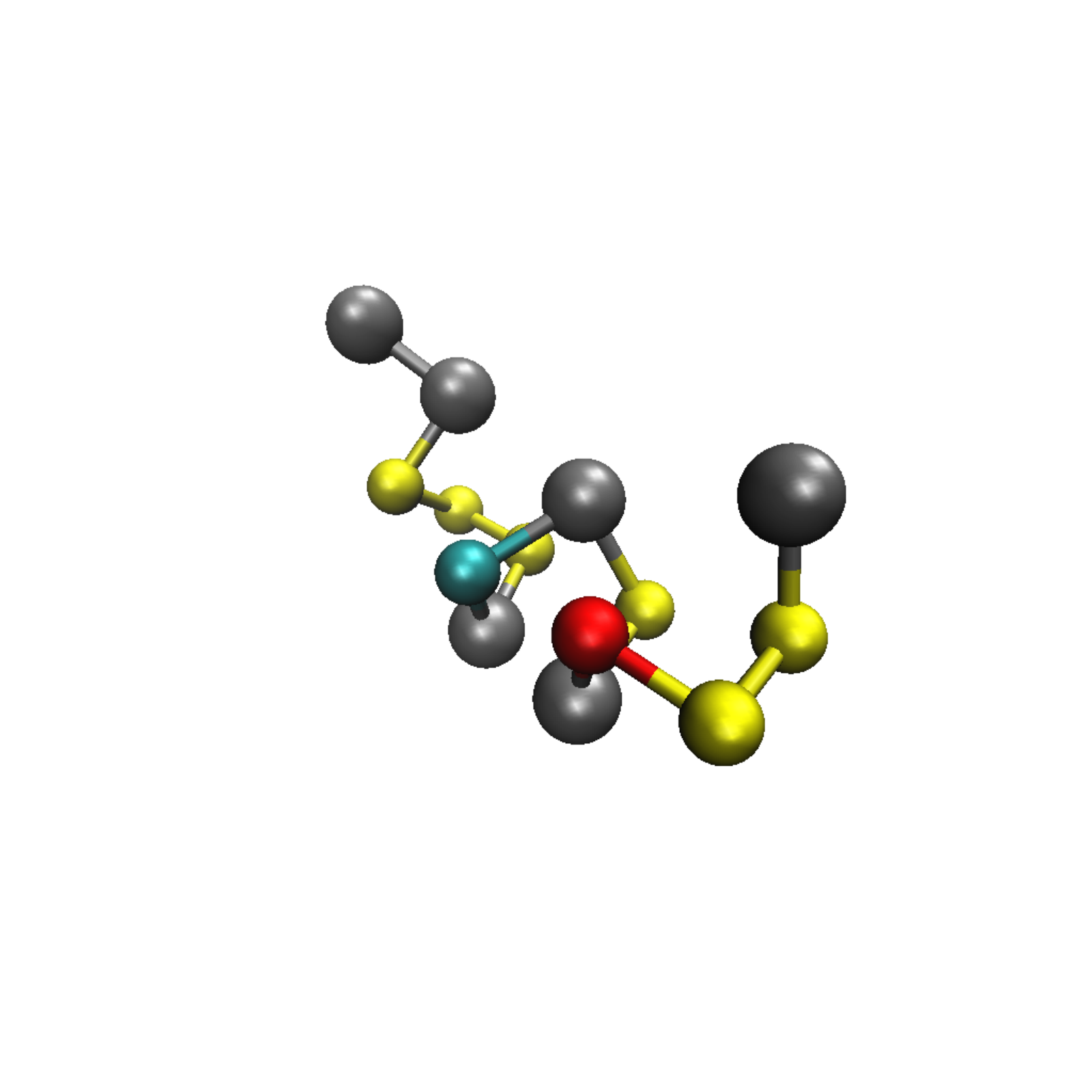}%
    }\hfill
    \subfloat[
        \label{fig:frustratedState}
        {State $29$ lacking a $[7, 11]$ bond}
    ]{%
        \includegraphics[width=0.3\textwidth]{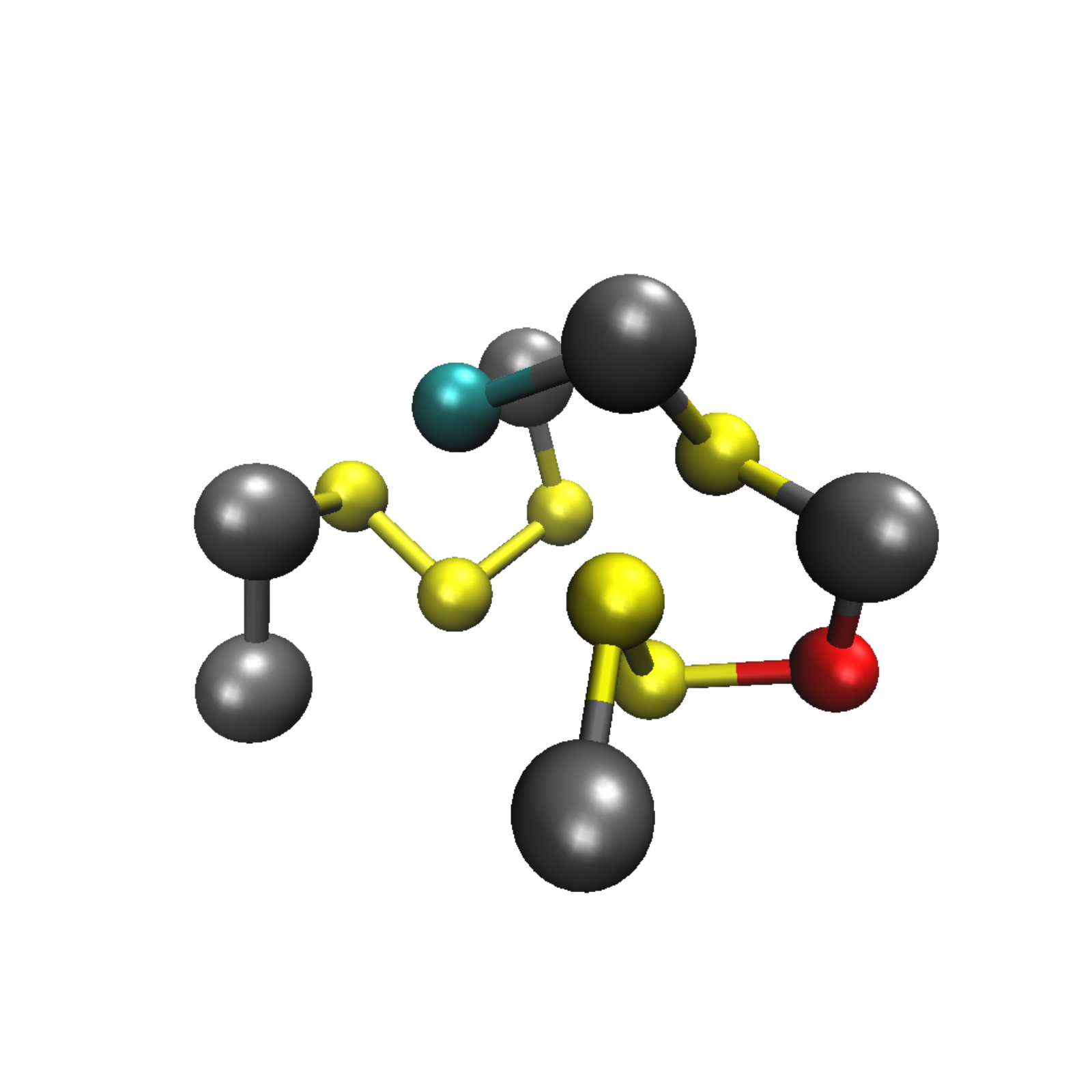}%
    }
    \caption{Example structures of the frustrated model system. On the left is
        the fully folded state (structure $37$), in which all the bonding
        constraints between bonding atoms (yellow) are satisfied. In the center
        is the transition state (structure $36$), satisfying all local helical
        bonds, with the largest flux between it and the final state. On the
        right is a kinetically trapped state (structure $29$), in which the bond
        between bead $7$ (blue) and bead $11$ (red) cannot be formed without
        breaking existing bonds. There are three low-lying trapping
    configurations in the model, identified as states $29$, $30$, and $34$.}
    \label{fig:frustratedModel}
\end{figure*}

\begin{figure}[htbp]
    {\centering{
    \includegraphics[height=6.5cm]{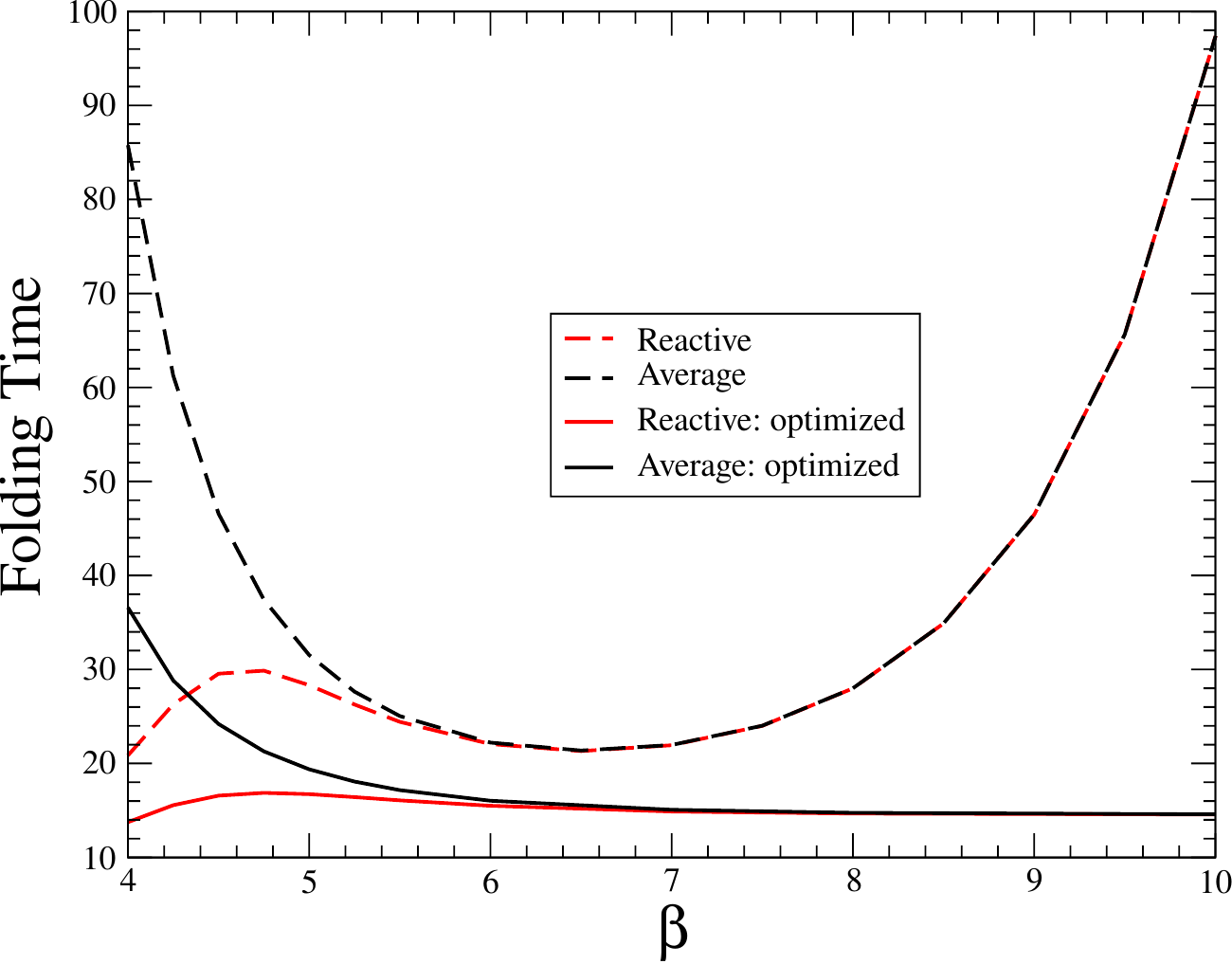}}}
    \caption{The reactive and average folding times (in units of $1 / D$) as a
        function of inverse temperature $\beta$ for the frustrated model. At
        high temperatures (low values of $\beta$), the equilibrium probability
        of the folded state is small, leading to a large folding time due to
        trajectories that return to the unfolded state. At low temperatures
    (large $\beta$), the slow transitions out of trapping states lead to folding
times that increase rapidly with $\beta$.}
    \label{fig:frustrateFoldingTimes}
\end{figure}

\begin{figure*}[htbp]
    \subfloat[
        \label{fig:frustratedTree}
        Frustrated model
    ]{%
        \includegraphics[width=0.45\textwidth]{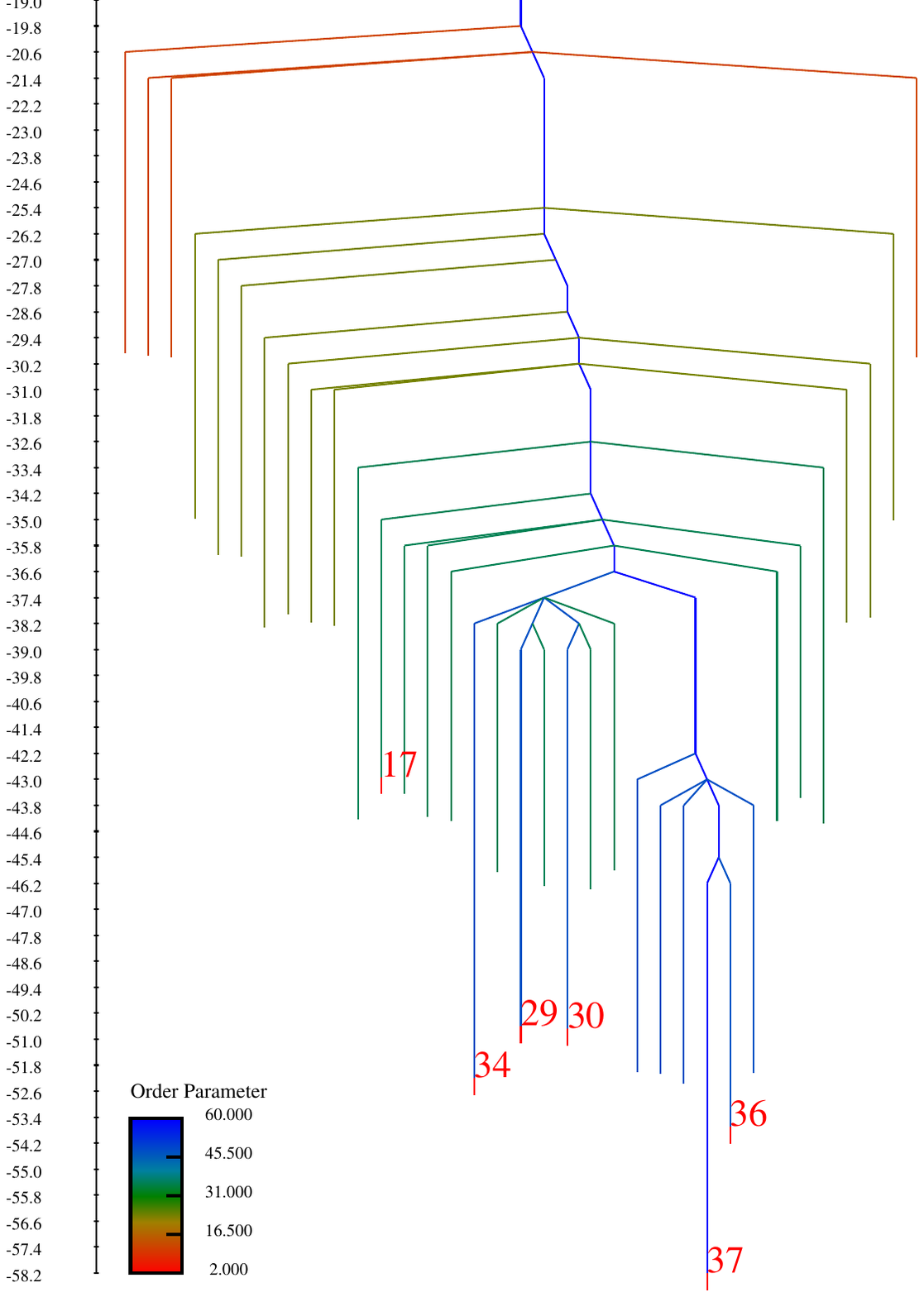}%
    }\hfill
    \subfloat[
        \label{fig:optimizedTree}
        Optimized model
    ]{%
        \includegraphics[width=0.45\textwidth]{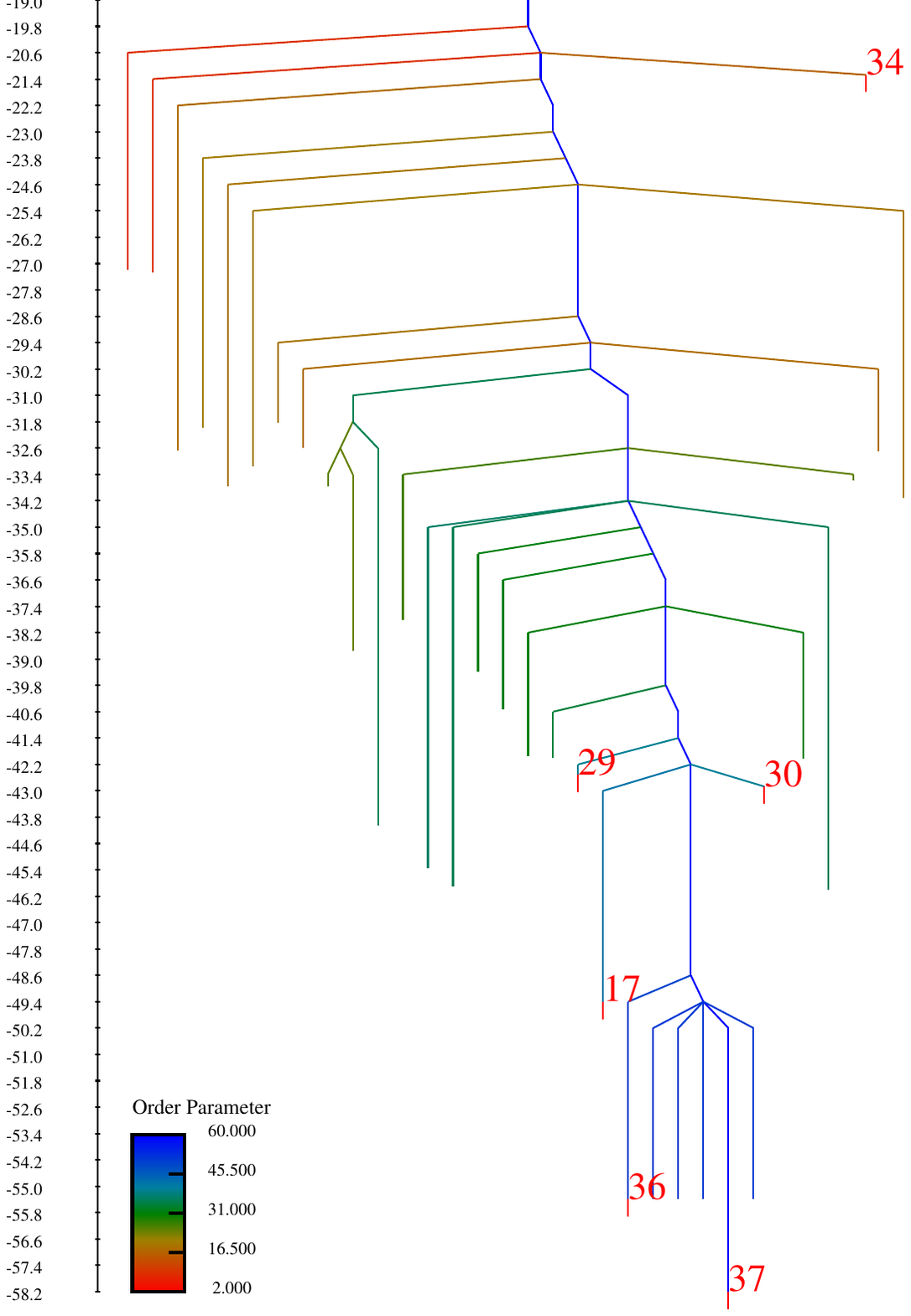}%
    }
    \caption{Disconnectivity graphs of the $14$-bead, $5$-bond model. Note the
    difference in locations of the trapping states $29$, $30$, and $34$ as well
as state $17$ in the respective graphs.}
\end{figure*}

\begin{figure}[htbp]
    {\centering{
    \includegraphics[height=5.5cm]{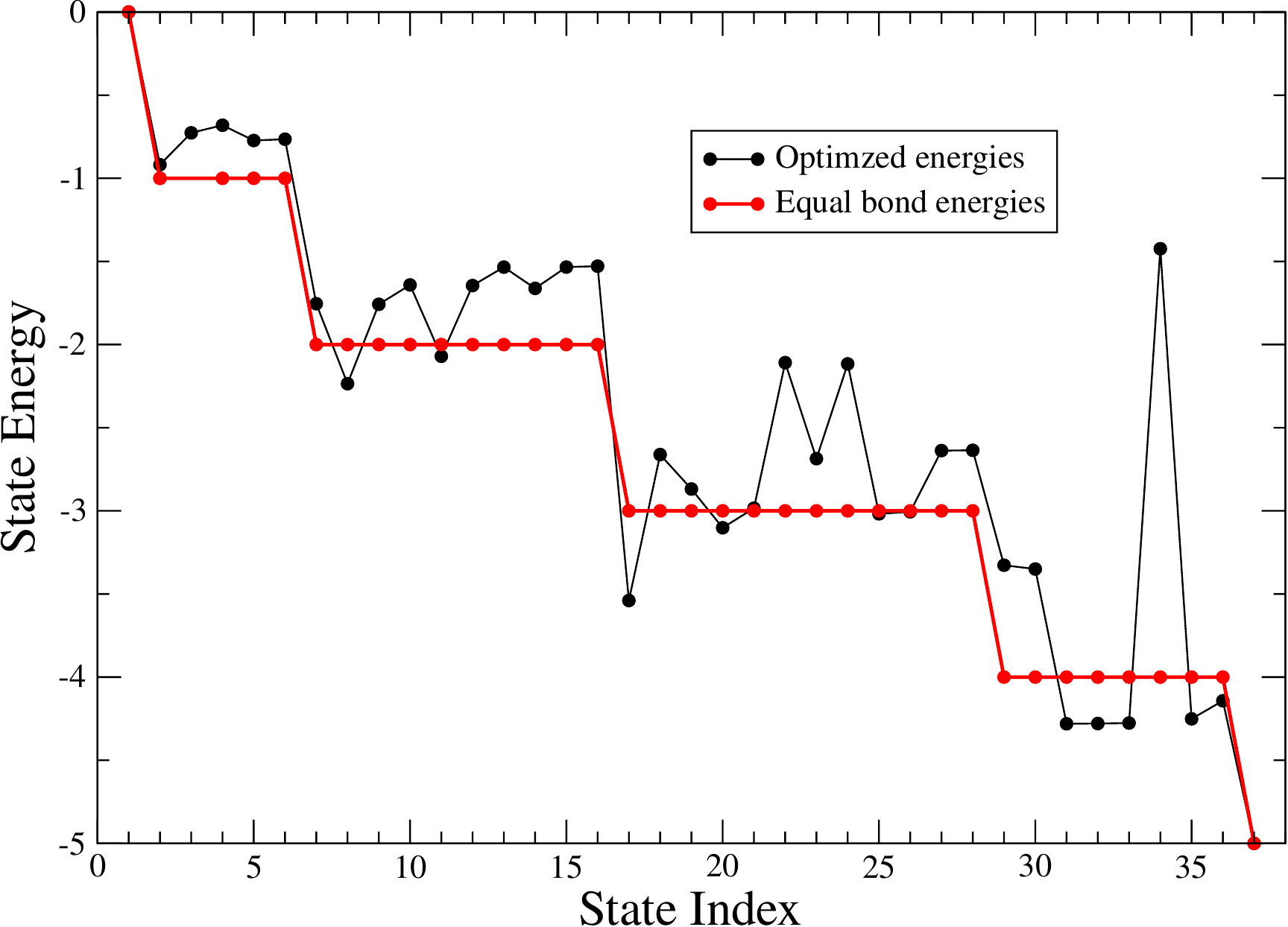}}}
    \caption{Energy adjustments to minimize the average folding time when the
        free energy difference between the folded and unfolded states is
        constrained to be $60$ and $P_u/P_f = 10^{-26}$. Note that the
        optimization raises the energies of the trapping states $29$, $30$, and
    $34$ to either minimize their path probability and flux, or facilitate
escape by lowering backward transition barriers to outlet state $17$ which now
lies below the energies of the trapping states.}
    \label{fig:optimizedEnergy}
\end{figure}

\begin{figure}[htbp]
    \centering
    \subfloat[
        \label{fig:frustratedNetworkDiagram}
        Network diagram: frustrated model
    ]{%
        \includegraphics[height=6cm, width=0.35\textwidth]{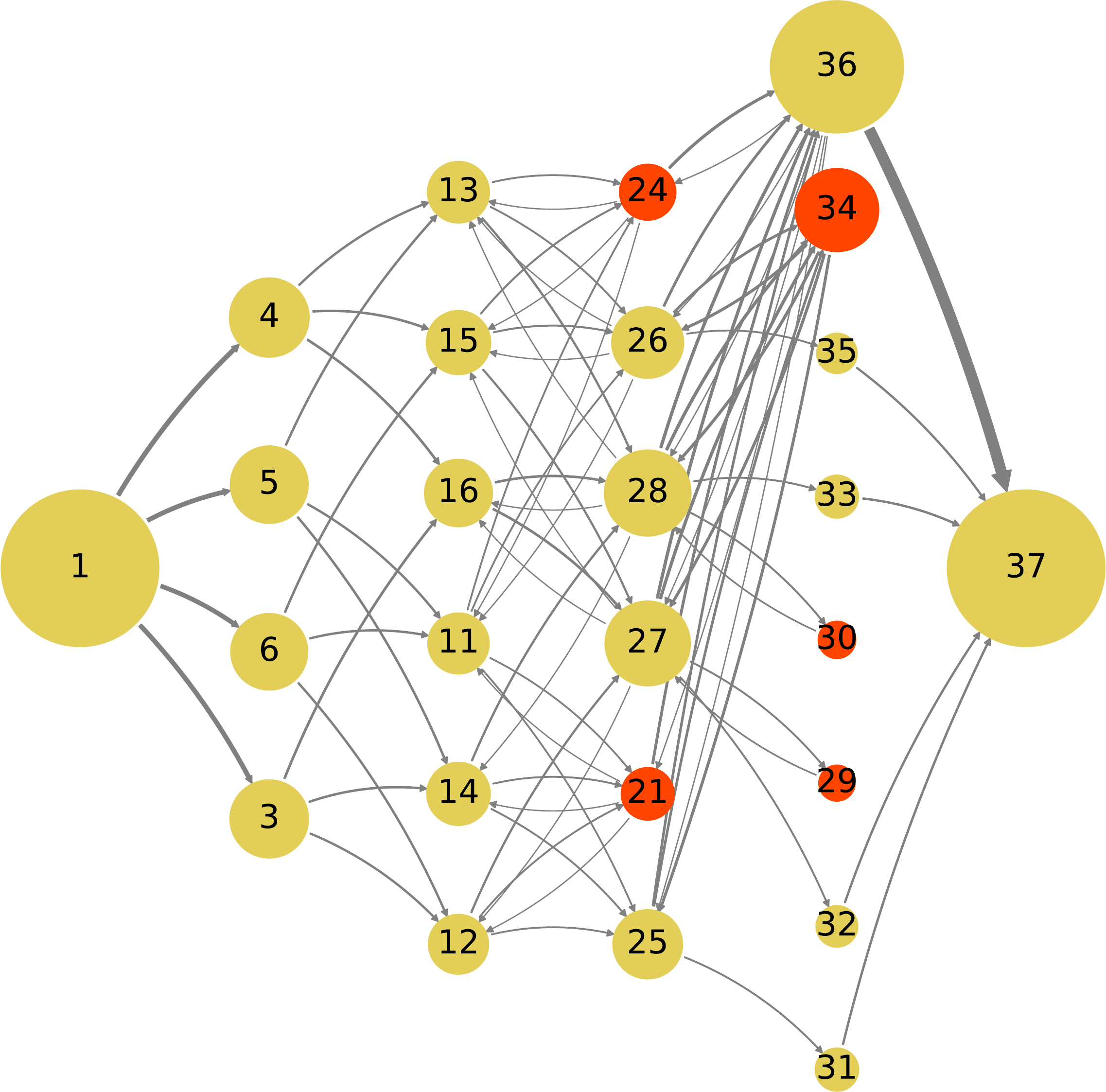}%
    }\hspace{20ex}
    \newline
    \subfloat[
        \label{fig:optimizedNetworkDiagram}
        Network diagram: optimized model
    ]{%
        \includegraphics[height=6cm, width=0.3\textwidth]{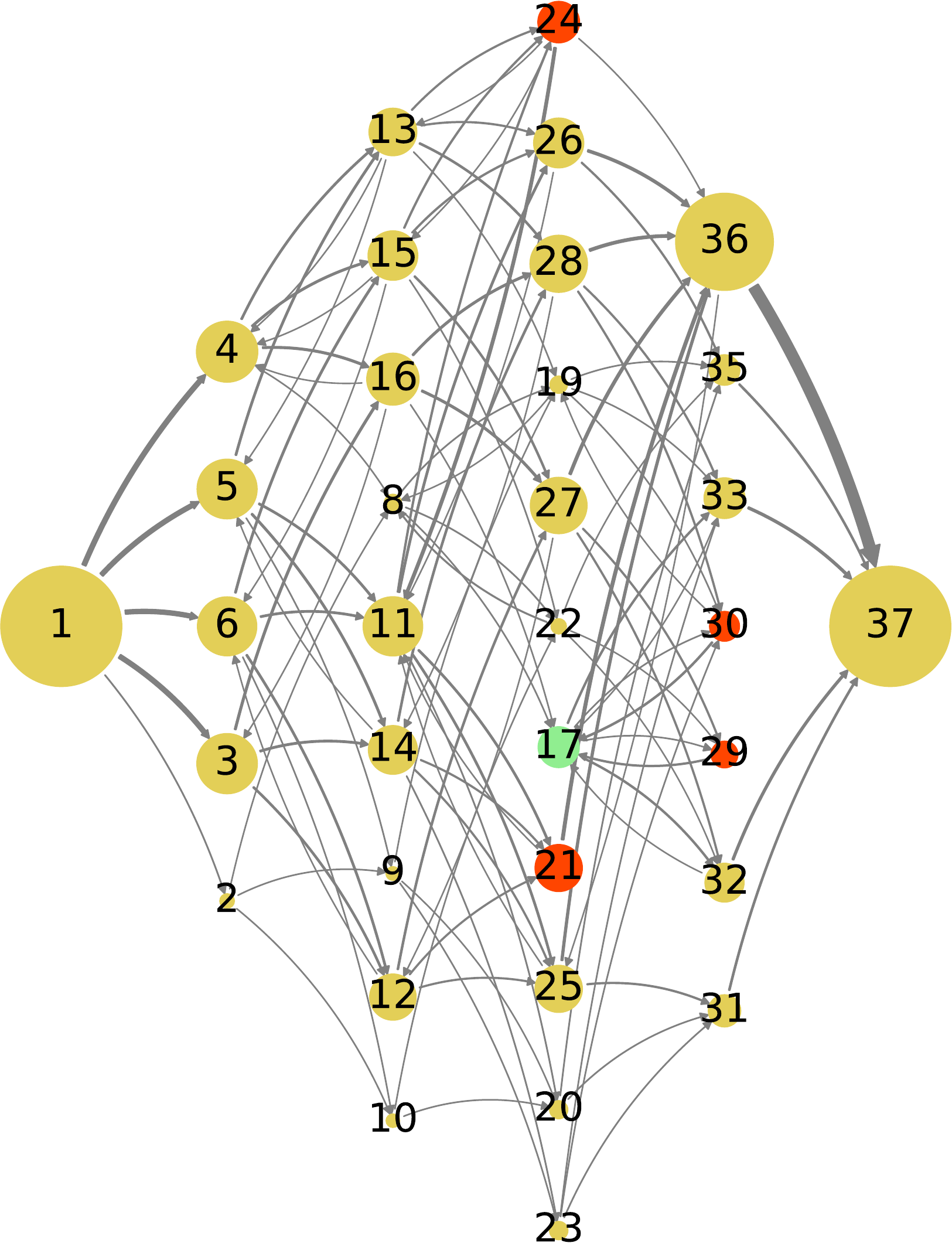}%
    }
    \caption{Network diagrams of the folding process for the frustrated model.
        The size of a node $i$ is representative of the probability $r^+_i$ that
        the state is visited in the reactive ensemble, and the size of the
        arrows between nodes represents the reactive flux between them (see
        Eq.~\eqref{eq:reactiveFlux}). On the left is the network for a system
        with bonds of equal energy, and on the right is the network following
        the optimization of the folding time. Note the disappearance of trapping
    state $34$ and the flux between trapping states $29$ and $30$ (all colored
red) and state $17$ (colored green) that appears in the optimized network.}
\end{figure}

\begin{figure*}[htbp]
    \subfloat[
        \label{fig:trappingFPT}
        Frustrated model
    ]{%
        \includegraphics[height=5.5cm]{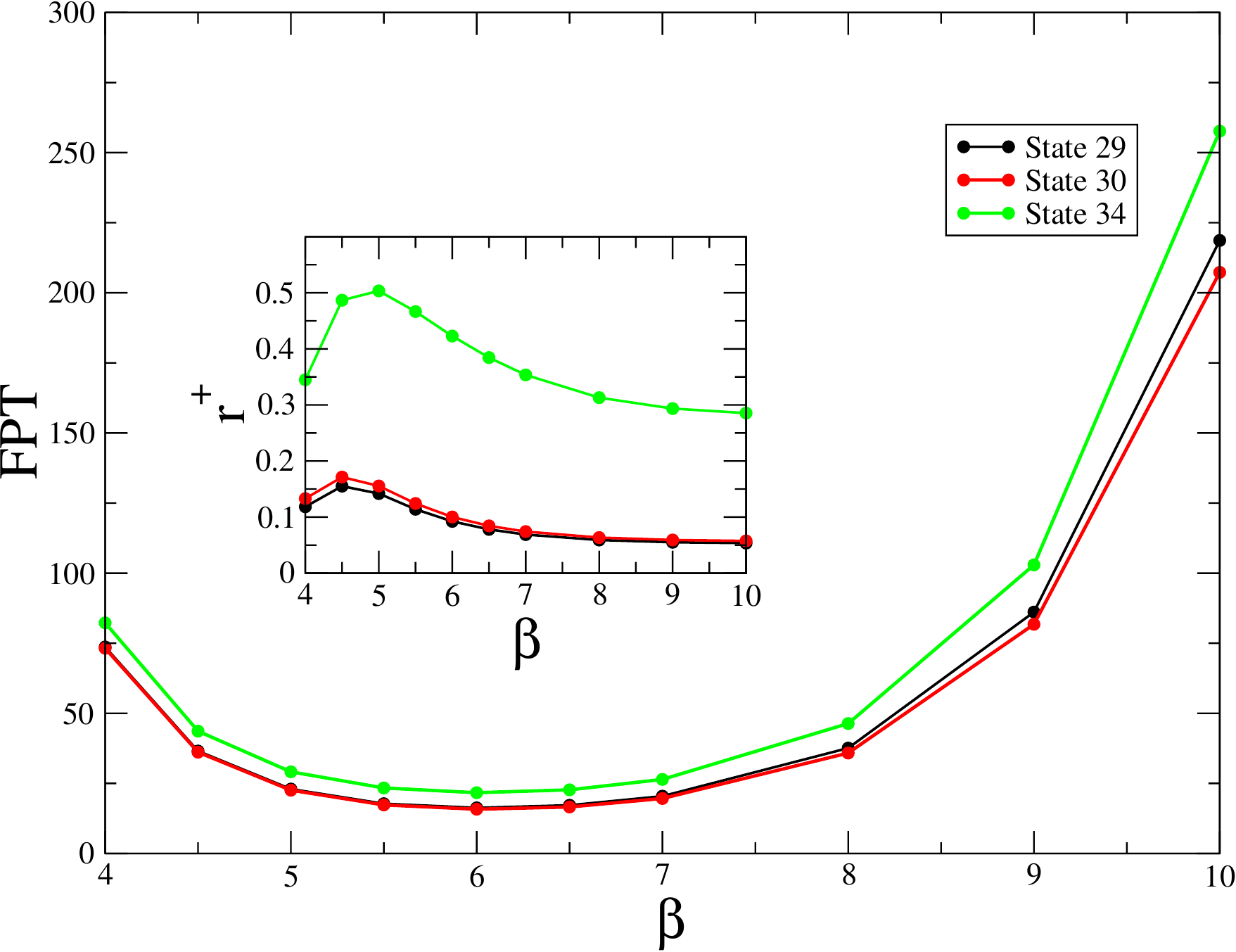}%
    }\hspace{15ex}
    \subfloat[
        \label{fig:optimizedFPT}
        Optimized model
    ]{%
        \includegraphics[height=5.5cm]{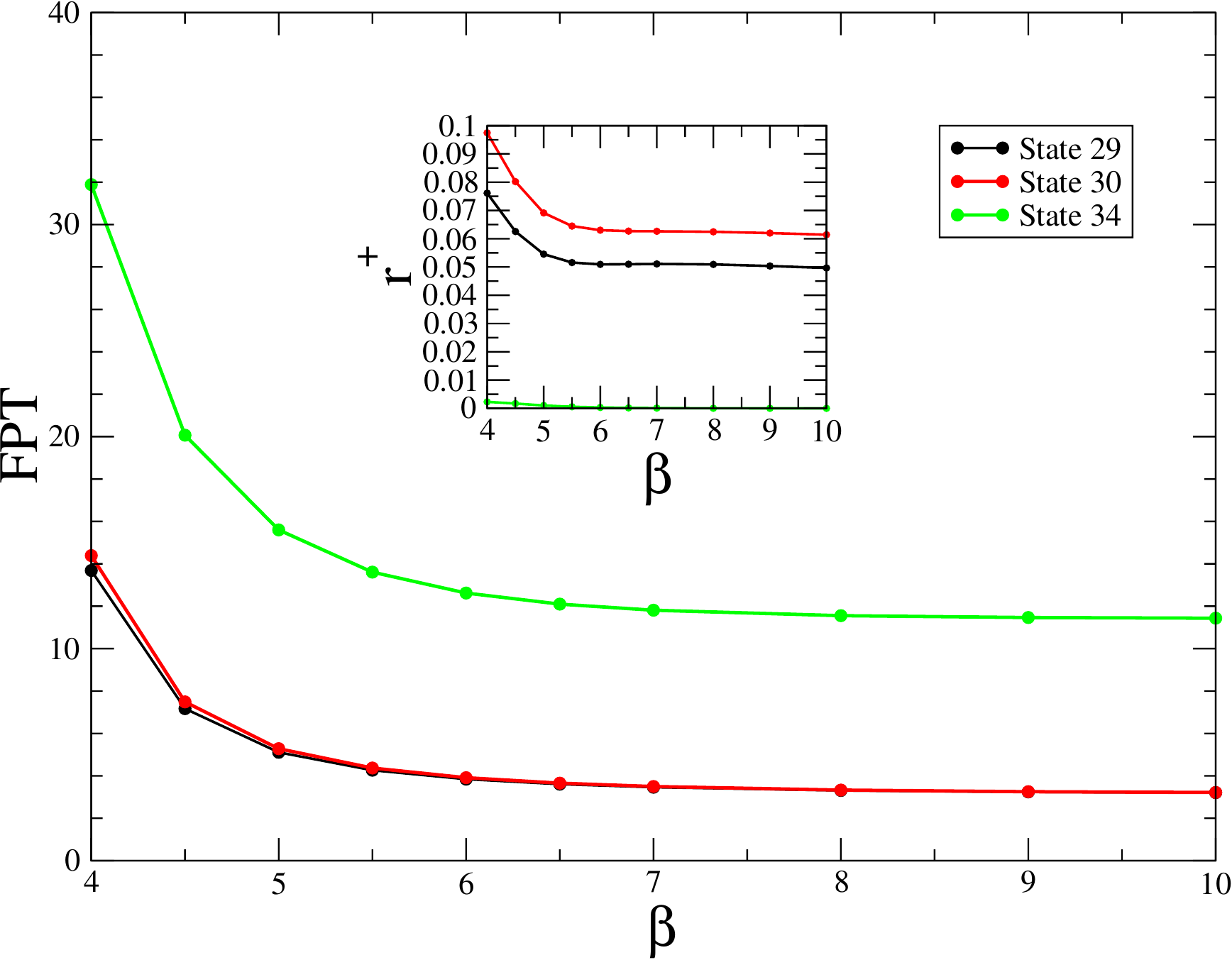}%
    }
    \caption{The first passage times in units of $1 / D$ from the trapping
        states to the folded state. The inset in both figures shows the reactive
        probability of passing through the trapping states. Note that the
        similarities of the first passage times to the mean first passage time
    in Fig.~\ref{fig:frustrateFoldingTimes} indicates that the first passage
times out of the three trapping states dictate the overall folding behavior of
both models.}
\end{figure*}

The multiplicity of competing interactions in real systems can give rise to free
energy landscapes with many local minima, resulting in long-lived metastable
structures. Small proteins that fold quickly have amino acid sequences that lead
to thermodynamically stable configurations and avoid kinetically trapped
metastable states.

To examine the role of kinetic traps and their elimination through a selection
process, we consider a $14$-bead model with a set of bonding interactions $\{[4,
12], [3, 7], [5, 9], [7, 11], [9, 13]\}$, all formed at $r_c = 1.5$.
Highly-bonded structures in this model, shown in Fig.~\ref{fig:frustratedModel},
resemble a short $\alpha$-helix that folds over due to the long-range
interaction between beads $4$ and $12$. The model mimics misfolding due to the
existence of kinetic traps: For the trapping states, the manner in which the
structure satisfies a set of bonding constraints geometrically prohibits the
formation of the additional bonds required to reach the fully bonded structure.
An example of such a structure is shown in Fig.~\ref{fig:frustratedState}. The
identification of trapping states and the calculation of their entropies can be
difficult. Within the layer approach, the trapping states are found by
identifying which and how many states in a pool of possible structures are
incapable of reaching a target state the next level down in a short trajectory.
Each of the pools is iteratively constructed in parallel, starting from the
fully unfolded state with no bonds. For a given state, its pool of structures is
generated by using the pools of all structures in the previous layer that can
reach the target state by the formation of a single bond.

For this model system, we find a total of five trapping states: Two of the
states are in the third level and have three bonds, and three are found in the
fourth level and cannot form the $[3, 7]$ bond (structure $29$), the $[5, 9]$
bond (structure $30$), or the $[9, 13]$ bond (structure $34$) due to the
preexisting long-range bond between beads $4$ and $12$. In
Fig.~\ref{fig:frustratedTree}, the trapping states appear as a separate fork in
the disconnectivity graph of the model system at low temperatures, since
dynamical events with high free energy barriers that break bonds must occur to
reach the fully folded state. The effect of the trapping states on the dynamics
is significant, and leads to a qualitatively different temperature dependence
from the fast-folding model of crambin. At high temperatures, as in the crambin
model, the fully folded structure is thermodynamically unfavorable and reactive
trajectories have low probability. Instead, the main contribution to the average
first passage time to the target state comes from the non-reactive paths that
repeatedly revisit the unfolded state. The probability of nonreactive
trajectories rapidly decreases with temperature, and the minimum folding time is
reached at intermediate values of $\beta$ near $\beta = 6$. At low temperatures,
many folding trajectories become kinetically trapped, and the folding time
increases exponentially as the free energy barrier increases. This kinetic
trapping, whose inverse temperature dependence is plotted in
Fig.~\ref{fig:trappingFPT}, follows the same trends as the mean folding time
(see Fig.~\ref{fig:frustrateFoldingTimes}).

As is evident in Fig.~\ref{fig:optimizedEnergy}, the constrained minimization of
the folding time with respect to the state energies eliminates the effect of the
trapping states by raising the energy of the trapping states so that either they
have negligible reactive probability $r^{+}$ at all temperatures (state $34$),
or they are in resonance with state $17$ in the previous level with fewer bonds
(states $29$ and $30$). In the optimized model, these states have a low
activation barrier and rapidly break the bond connecting them to the less bonded
state $17$, which appears with enhanced reactive probability. At the same time,
the flux of non-trapping states in the final layer is optimized by lowering the
energies of those states to allow multiple pathways of similar probability to
pass to the target state. These effects are evident in the disconnectivity graph
of the optimized model in Fig.~\ref{fig:optimizedTree} by the equal barrier
heights of the layer of states and the shift of trapping states $29$, $30$, and
$34$ to higher points. Additionally, the trap outlet state $17$, which contains
the $[4, 12]$, $[7, 11]$, and $[9, 13]$ bonds, is repositioned in the tree-like
structure.

The changes in the network diagrams of the folding (see
Fig.~\ref{fig:optimizedNetworkDiagram}) highlight the disappearance of node $34$
and the increased flux of transitions among states in the middle levels of the
network. In both the equal bond energy and optimized energy models, the main
final transition occurring in $70\%$ of the folding trajectories to the target
state consists of the formation of the $[4, 12]$ bond. As a result, the
qualitative nature of the folding pathways is similar to that of the crambin
model: The system first forms a helical element that subsequently folds into the
final structure. After optimization and removal of the trapping kinetics, the
temperature dependence of the mean first passage time from the unfolded to
folded states in Fig.~\ref{fig:optimizedFPT} approaches a constant as $\beta$
increases, the same qualitative behavior observed in the rapid-folding crambin
system.

\section{Discussion and conclusions} \label{sec:summary}

In this work, we introduced methodology to address the computational challenges
of computing the entropy and mean first passage times for a linear chain model
of proteins in which monomers interact discontinuously. These quantities appear
as parameters in Markov state models of the population dynamics. The methods
combine adaptive sampling algorithms with statistical tests to compute reliable
interval estimates for all quantities. Given the exponential growth of the
number of states with the number of bonding interactions included in the model,
parallel algorithms are a critical requirement to investigate large, complex
models. The level-based calculations in which individual pairs of linked states
are conducted in parallel with the inclusion of either replica-exchange or
population Monte Carlo components improve the rate of convergence for the entire
set of state entropies and allow the computation to be carried out in massively
parallel platforms with coarse-grained parallelism. The numerical sampling
difficulties associated with large first passage times are lessened by the
introduction of intermediate staircase states, which was shown to significantly
reduce the sampling required for a given statistical resolution.

There are possible improvements to the sampling that are relatively simple to
implement. The methods presented here rely on sampling states using a model with
a discontinuous potential with event-driven dynamics that prevent the
implementation of continuous adaptive biases frequently used in the molecular
simulation community. However, auxiliary sampling chains based on dynamical
trajectories governed by continuous potential approximations to Heaviside and
infinite square well functions can be applied to generate trial Monte Carlo
updates, provided that the acceptance criterion is suitably
adjusted\cite{Radu:2000}. The continuous potentials can be adaptively adjusted
along bonding distances using methods such as well-tempered
metadynamics\cite{Laio:2002,Barducci:2008}, and self-healing umbrella
sampling\cite{Marsili:2006}. However, some care is required to ensure that the
continuous potential system does not frequently allow configurations that
violate the strict geometrical constraints of the model. Current studies along
these lines are underway.

Given the computational cost of models with many nonlocal bonds, an interesting
open question is whether or not the methods of machine learning on small systems
can be used to accurately infer the entropies and first passage times of more
complicated models. Nonetheless, machine learning algorithms frequently require
large sets of training data to be useful. The sampling algorithms introduced
here can help with the task of generating the necessary training data sets.

Machine learning methods may also prove useful in the classification of trapping
states that subdivide configurations determined by their bonding patterns alone.
Geometrical descriptors that generalize the state indicator functions will allow
for more accurate evaluation of the probability of the kinetic traps as well as
their first passage times.

The main appeal of the Markov state model of the discontinuous potential lies in
the possibility to evaluate both the structure and dynamics for an infinite
number of choices of interactions at arbitrary temperatures, once the density of
states and the first passage times have been calculated for a given choice of
bonding pattern. In Secs.~\ref{sec:crambin} and \ref{sec:frustratedModel}, we
demonstrated how this flexibility may be exploited to select interaction
energies that enhance physical properties or desired functional characteristics.
In these sections, we analyzed the mean passage time from the unbonded state to
the fully bonded state for a model of the crambin protein, which folds quickly
and has a free energy landscape with a funnel-shaped morphology, and for a short
$14$-bead helical protein designed to exhibit a more complex free energy
landscape and trapping kinetics.

For the crambin model, a choice of pairwise-additive bond energies for states
led to a simple mechanistic folding pathway, in which the helical portions of
the model protein formed first with no clear preference of order, followed by
the passage with near unit probability through a helical bottleneck state. In
the second step of the folding, distant bonds linking regions of the helix to
one another lead to a penultimate state with all bonds present except for those
linking the most distant edges of the chain. The optimization of the state
energies with a fixed relative probability of the unfolded to folded states
resulted in a different folding mechanism and folding rates that were twice as
large. Interestingly, the initial stage of the folding process in the
optimization involved the rapid formation of the local helical bonds. However,
the passage through the restrictive bottleneck state was discouraged by
adjustments that lowered the energy of states with long-range bonds. Similarly,
multiple pathways to the final state were found due to energy adjustments of the
long-range bonds to compensate for their different entropic values.

The folding process for the $14$-bead model system with frustration was also
initiated by the formation of local interactions and a helical precursor to the
final folded state. However, the inclusion of a high density of interactions in
the model introduces a number of kinetic traps that are reflected in an
exponentially increasing folding time as the temperature decreases, and new
distinct branches appear in the disconnectivity graphs of the free energy
landscape at low temperatures. The deepest-lying trapping states determine the
folding time at low temperatures. In this case, the optimization of energies
destabilized the trapping states so that they either had a negligible
probability in the folding pathway or were positioned in resonance with states
with fewer bonds to enable the rapid breaking of a bond.

The optimal energies for rapid folding depend on the choice of constraints
employed in the optimization procedure, and these constraints should reflect
conditions that are realizable for molecular systems. If the bonds formed are
intended to represent weak electrostatic or hydrogen bonding interactions
between segments of the chain, the maximum drop in the state's energy should be
restricted in magnitude. Large increases in the state energies can easily be
achieved through steric repulsion or torsional strain. It is likely chain
stiffness along the peptide backbone effectively limits the density of bonds in
the chain to avoid this type of kinetic trap. It would be interesting to explore
the inclusion of other information in the loss function. For example, a target
electrostatic map for the folded structure could be included as a penalty in the
loss function and monomer-specific partial charges used to determine optimal
residue sequences.

The linear chain model can be generalized to include side chain beads
interacting with other beads to allow for the inclusion of steric effects of
bulky residues as well as attractive nonlocal bonding. Such features are
important in determining the overall three-dimensional structure of real
proteins. The sampling methods and optimization procedure of the Markov state
model introduced here can be applied without modification.

The folding mechanism and optimization of the state energies in the models of
fast-folding proteins analyzed in this work are indicative of the type of issues
that can be explored with the discontinuous potential model. Its simplicity
opens the door to explore general questions that are difficult to address by
other means. Other avenues to be explored include the following: Given a
particular three-dimensional structure, what state energies lead to rapid
folding and thermodynamic stability? To what extent is the optimization of the
native state of a protein for fast refolding dictated by its topology? How
additive are the energies of biopolymers? How do biomolecular systems avoid
kinetic traps? Why do certain motifs of secondary structure appear and not
others, and what role do secondary structures play in the folding pathways? Does
the optimization of the folding time confirm well-established principles of fast
folding proteins, such as the importance of Ramachandran angles, the existence
of foldons and the statistical correlation between contact
order\cite{Plaxco:1998,Dinner:2001} and folding rate? What are the differences
in the interaction patterns of fast-folding vs intrinsically disordered
proteins? Work along these lines is underway.

\section{Acknowledgements}

Financial support from the Natural Sciences and Engineering Research Council of
Canada is gratefully acknowledged. Computations were performed on the Cedar
supercomputer at Simon Fraser University, which is funded by the Canada
Foundation for Innovation under the auspices of Compute Canada, WestGrid, and
Simon Fraser University. Code for this project is available at:
\url{https://github.com/margaritacolberg/hybridmc}.

\bibliography{mfpt-sampling}

\end{document}